\documentclass[showpacs,aps,prl,preprint,superscriptaddress]{revtex4}
\usepackage{graphicx,epsf}
\begin{document}

\title{First-principles molecular-dynamics simulations of a hydrous silica melt:
Structural properties and hydrogen diffusion mechanism}
\author{Markus P\"ohlmann$^{1,2,3}$, Magali Benoit$^1$ and Walter Kob$^1$\\
\small{\it 1 - Laboratoire des Verres, Universit\'e Montpellier II, 34095 Montpellier, France}\\
\small{\it 2 - Physik-Department E13, Technische Universit\"at M\"unchen, 85747 Garching, Germany}\\
\small {\it 3 - Inistitut Laue-Langevin, 38042 Grenoble, France}
}
\begin{abstract}
We use {\it ab initio} molecular dynamics simulations to study a sample 
of liquid silica containing 3.84 wt.$\%$ H$_2$O. 
We find that, for temperatures of 3000 K and 3500 K,
water is almost exclusively  dissolved as hydroxyl groups, the silica network is partially broken 
and static and dynamical properties of the silica network change considerably upon the addition of water. 
Water molecules or free O-H groups occur only at the highest temperature but are not stable and disintegrate rapidly.
Structural properties of this system are compared to those of 
 pure silica and sodium tetrasilicate melts at equivalent temperatures. These comparisons confirm the picture 
of a partially broken tetrahedral network in the hydrous liquid 
and suggest that the structure of the matrix is as much changed by the addition of water than it is 
by the addition of the same amount (in mole \%) of sodium oxide. 
On larger length scales, correlations are qualitatively similar but seem to be more pronounced
 in the hydrous silica liquid.
 Finally, we study the diffusion mechanisms of the hydrogen atoms in the melt.
 It turns out that HOSi$_2$ triclusters and SiO dangling bonds play a decisive role as intermediate states
 for the hydrogen diffusion.
\end{abstract}

\maketitle
\section{Introduction}
Over the last decades, amorphous silicates containing several wt.$\%$ of water have attracted the interest of many 
experimental and theoretical research groups due to its importance in science and technology. E.g. these systems are supposed to play an important role in 
magmatic flow in the earth crust and have therefore been investigated 
extensively by many geologists (for reviews see \cite{Ko00,MM94}). In silicate melts (lava), water can be released if temperature and pressure conditions are altered and in that case, the released water molecules assemble in bubbles. 
 These bubbles can set up pressure for explosive volcanism \cite{Di96, Sa99}. 
The microscopic origin of such bubbles is still very poorly understood \cite{Sa99} and one of the main motivating aspects for our study is to find indicators for such bubble formation. 
The possible use of water containing silica in 
fuel cells \cite{No97,No99} and the use of water as defect passivant in semiconductor 
devices (and reladed problems) \cite{He94} and its disturbing presence in optic fibers \cite{Fr86} underline the technological importance.  \\  
Over the past years pure amorphous silica has been well understood regarding its electronic 
and structural properties and experimental and simulational data seem to agree well 
\cite{Sa95,Sa95a,Pa97,Be00,Su91}. In recent studies these investigations have been extended 
to the technologically highly important class of sodium silicates  \cite{Is01,Is02,Ho01,Ho02,Zo95,Sm95,Ov98}. 
New fundamental ideas of the structural properties and the diffusion mechanism of sodium atoms in these melts 
have been obtained. \\
The impact of water on the structure and dynamics of silica is supposed to be at least as important as the addition of sodium \cite{Ba86}. Various methods have been used to investigate water speciation and diffusion in a SiO$_2$ network. NMR, Infrared and Raman spectroscopy as well as neutron scattering experiments 
suggest a chemical reaction of molecular water to SiOH groups in the melt for low initial water 
concentrations and a saturation of this reaction if more water is added to the liquid 
\cite{Ko00}. In chemical terms it represents a balance of the form
\begin{displaymath}
{\rm Si-O-Si+H_2O} \longleftrightarrow {\rm 2 (SiOH) } 
\end{displaymath}
that follows the  Chatelier principle  \cite{Ho95} and which can be shifted to any side by the variation of 
external conditions like temperature or concentration of one species. The energetics 
is supposed to be such that a reaction to the right is endothermal and hence at high temperatures 
SiOH is the dominant species \cite{Ko00}. 
On the other hand, the mentioned experiments have difficulties to provide reliable quantitative data. The different fabrication processes of the samples and the overlapping of hydroxyl and water signals in the mentioned spectra are the main reasons for this \cite{Ko00,Be03}. In-situ measurements in the geologically relevant liquid state are hardly accessible due to elevated temperatures and pressures. \\
Computer simulations do usually not encounter these problems since, in principle, they can be done at high temperature
and pressure. 
Nevertheless problems for computer simulations that use  inter atomic
potentials arise directly from the above reaction: Potentials that describe appropriately all possible 
types of oxygen (bridging oxygens in the tetrahedral network, oxygens in hydroxyl groups, molecular 
water oxygens) have been developed with some success but do not describe the dynamics very well 
\cite{Fe90,We98}. Here the great advantage of {\it ab initio} simulations comes into play as the 
potential and the speciation are found and updated from accurate electronic structure calculations  
at each molecular dynamics time step. Especially {\it ab initio} Car-Parrinello simulations turned 
out to be a powerful tool for investigations of disordered silica and sodium silicate systems 
\cite{Sa95,Sa95a,Pa97,Is01,Is02,Be02}. For a review of previous {\it ab initio} simulated amorphous 
systems see \cite{Pa01}. \\
In this paper we present an equilibrated liquid of water-containing SiO$_2$ and its structural 
and dynamical properties. These properties will be compared to that of a silica and a sodium silicate
melt when possible. 
The paper is organized as follows: In section \ref{sec:simdet} we discuss the used simulation techniques and in section \ref{sec:res} we present the structure of the liquid and draw some conclusion on diffusion mechanisms for hydrogen. Conclusions are summarized in the last section.  

\section{Simulation Details}\label{sec:simdet}
The molecular dynamics simulations were performed with the Car-Parrinello code CPMD \cite{Ca85,Hu99} at 3000 K and 3500 K. 
The electronic subsystem was treated with a density functional (DFT) approach in a generalized gradient 
approximation using the PBE functional \cite{Ko65,Pe96}. The core electrons of silicon and 
oxygen atoms were treated in a pseudo potential model using Troullier-Martins type pseudo potentials \cite{Tr91}. 
A plane wave $\Gamma$ point expansion with an energy cutoff of 50 Ry turned out to be sufficient 
for an appropriate description of the inter atomic forces. 
The 50 Ry cutoff with the PBE functional was first tested on the H$_2$O dimer and on $\alpha$-quartz. 
These tests showed that e.g. the two SiO distances of $\alpha$-quartz are equal to 1.624 \AA\  and  to 1.628 \AA\ 
 independent of the energy cutoff between 50 Ry and 100 Ry. On the other hand, we found for the H$_2$O dimer that 
the O-O distance, the quantity which is the most sensitive to a change of the cutoff, shows only a variation from 2.925 \AA\
 to 2.950 \AA\  if the cutoff is decreased from 90 Ry to 50 Ry. \\
For the SiO$_2$-H$_2$O liquid we used a simulation box containing 102 atoms (30 silicon, 64 oxygen, 8 hydrogen) 
corresponding to a concentration of water of 3.84 wt.\%. The simulation box was prepared by starting from an equilibrated 
liquid silica configuration (obtained with the effective potential proposed by van Beest {\it et al} \cite{BKS}) 
with 32 SiO$_2$ units. Two randomly chosen silicon atoms were then  substituted by four hydrogen atoms each so that 
the chemical stoichiometry was preserved. These hydrogen atoms were attached to the oxygen atoms which were formerly bound 
to the removed silicon atoms. Using CPMD the liquid was subsequently equilibrated at 5000 K for some time and finally 
cooled down to 3500 K and 3000 K and annealed for roughly 5 ps before recording the trajectories for the present work.   
Since, to our knowledge, densities of water-containing silica systems 
(especially in the liquid state) have not been measured, the volume of the box had to be 
chosen such that the calculated internal stress of the system had a mean value of zero. This was the case 
for a box  size of 11.50 \AA\ and hence a density of 2.05 g/cm$^3$. 

Comparisons of structural quantities like pair distribution functions, mean square displacements or 
angle distributions extracted from molecular dynamics simulations performed at 3500 K with 
 a cutoff of 50 Ry and of 80 Ry did not show differences within the statistical error bars. 
Thus we concluded that the 50 Ry cutoff was sufficiently high. 
For equilibration, the masses of the ions were all set to 28 a.u. (the mass of a silicon atom). 
Note that a change of the  ionic masses does not affect the structure of the liquid since 
 at equilibrium all structural quantities are independent of the mass. 
On the other hand, the increase of the ionic masses (from 1 to 28 for hydrogen and from 16 to 28 for oxygen) 
should increase the energy gap in the electronic structure and allow an increase of the Car-Parrinello electronic mass 
and hence the use of a larger time step which thus leads to a faster equilibration. 
The equilibration of the system was performed at the two ionic temperatures of 3000 K and 3500 K 
employing Nose-Hoover thermostats and an electronic mass of 600 a.u. 
(energy $\times$ time$^2$) at a time step of 4.5 a.u. (0.1088 fs). At high temperature,
the electronic gap is too small compared to $k_{\rm B}T$
to ensure the decoupling of the ionic and the electronic 
degrees of freedom, which is needed to perform Car-Parrinello dynamics. The use of thermostats 
is therefore compulsory. To speed up the equilibration and 
to perform an efficient canonical sampling, one separate Nose-Hoover thermostat chain for each 
ionic degree of freedom was used (known as ``massive'' thermostating \cite{Ma96}). 
The electrons were controlled with one single thermostat chain \cite{Ma92,Tu94}.  
Unfortunately, due to the use of thermostats the direct access to dynamical properties is no longer available. \\
\begin{figure}[ht]
\centerline{\includegraphics[width=6cm]{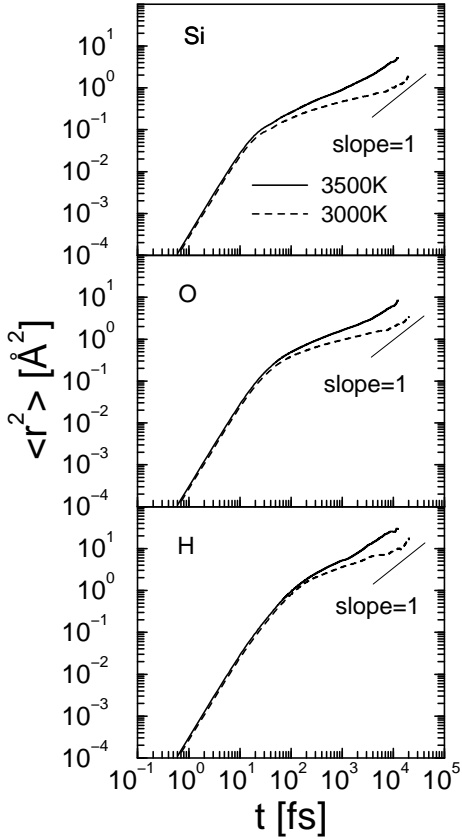}    }
\caption{\label{fig:msd} Mean square displacements of the Si, 
O and H atoms at 3500 K (solid lines) and 3000 K (dashed lines).}
\end{figure}

The system was equilibrated at two temperatures (3500 K and 3000 K) until in a log-log plot the averaged mean square displacements (MSD)
of the Si atoms showed at long times a slope close to unity. This is illustrated in Fig. \protect\ref{fig:msd}
where the mean square displacements (MSD) of the different atomic species are shown for the 3000 K
(dashed lines) and the 3500 K (bold lines) liquids. Usually MSDs of viscous liquids are composed of three regions: 
The ballistic one in which the atoms move without noticing their neighbors and hence a MSD that is proportional to $t^2$. This ballistic region is followed by a region where the atoms are temporarily confined 
in a cage made of their nearest neighbors. In this regime, the atoms rattle around in the cage without significant 
displacement, leading to a MSD that increases only slowly. Finally the atoms leave this cage and start to show a diffusion motion, i.e. a MSD that is proportional to $t$. 
The choice of the masses and the thermostats affect also the MSD. However, the height of the plateau and 
the displacement at the onset of the diffusional regime should be independent of the thermostat. 
Hence,  we consider the system to be equilibrated once the diffusional regime is reached which was the case after 4.4 ps 
at 3500 K and 10.9 ps at 3000 K. 

In order to check that the liquids were indeed  well equilibrated and that there were no aging effects, the
trajectories were cut into three equal parts.  The averaged mean square displacements were then calculated for 
each part separately and compared to each other. Since the three different averaged MSD did not show 
any drift, aging effects can be excluded and equilibration was indeed obtained after the above mentioned times. 
All structural quantities were extracted from the beginning of the  
diffusional part up to the total length of the trajectory (4.4 to 12.5 ps at 3500 K and 10.9 to 22.5 ps at 3000 K). 
From Fig. \ref{fig:msd} we note that at the end of the 3500 K trajectory the MSD is 29.2 \AA $^2$, 
8.1 \AA $^2$, and 5.0 \AA $^2$  for the H, O and Si atoms respectively. At 3000 K the MSDs were 16.4 \AA $^2$, 
3.5 \AA $^2$ and 1.73 \AA $^2$. 
According to the above mentioned equilibration times and the time step of 4.5 a.u. (0.1088 fs), 
the numbers of computed time steps were 114900 at 3500 K and 206800 at 3000 K. 
Using a Hitachi SR8000 computer, one time step takes 56 s on a single processor.  \\

\section{Results and Discussion}\label{sec:res}
In this section we present the structural analysis of the liquids at 3000 K and 3500~K. 
Commonly considered quantities like pair distribution functions, bond angle distributions,
coordination numbers, $Q$-species distributions and bridging to non-bridging oxygen ratio will be discussed and,
 as far as possible,  be compared to the data extracted from {\it ab initio} simulated silica 
and sodium silicate (NS4) melts \cite{Is02,Be01}. In particular the comparison to a recently investigated 
sodium tetrasilicate melt \cite{Is02} is highly interesting since  the  valence shell configuration 
of the sodium and the hydrogen atoms are equivalent.\\
In order to analyze the structure with particular attention to the formation and the rupture of the silica network, 
water molecules and hydroxyl groups, it turns out to be useful to distinguish several types of oxygen atoms: \\

\begin{tabular}{ll}
O$^*$ & hydroxyl group oxygen Si-O-H\\
 & (oxygen with one hydrogen and one silicon nearest neighbor)\\
wO & water molecule oxygen  H-O-H \\
 & (oxygen with at least two hydrogen atoms as nearest neighbors) \\ 
BO & bridging oxygen Si-O-Si \\
 & (oxygen with two silicon atoms as nearest neighbors)\\
NBO & non-bridging oxygen  Si-O(-?) \\
 & (oxygen with less than two silicon nearest neighbors)\\
O3  & tricluster oxygen  \\
 & (oxygen with three nearest neighbors) \\
O3H  & hydrogen containing tricluster \raisebox{1.1ex}{Si}\hspace{-0.35cm}\raisebox{-1.1ex}{Si} $>$O-H \\
    & (oxygen with two silicon and one hydrogen neighbor) \\
\end{tabular} \\ 

Note that, within these definitions, all the O$^*$ are also counted as NBO and that 
all the O3H are also counted as O3 as well as BO. The wO oxygen atoms can also
have silicon neighbors, e.g. the oxygen atom in H$_2$OSi is considered  as a wO. \\
The Si and H nearest neighbors of the oxygen atoms were determined as the Si and H atoms being located 
within a sphere the radius of which is given by the positions of the first minima in the Si-O and O-H
 pair distribution functions, respectively (see section \protect\ref{sec:RDF}).

\subsection{Radial Distribution Functions}
\label{sec:RDF}
\begin{figure}[ht]
\centerline{\includegraphics[width=5cm]{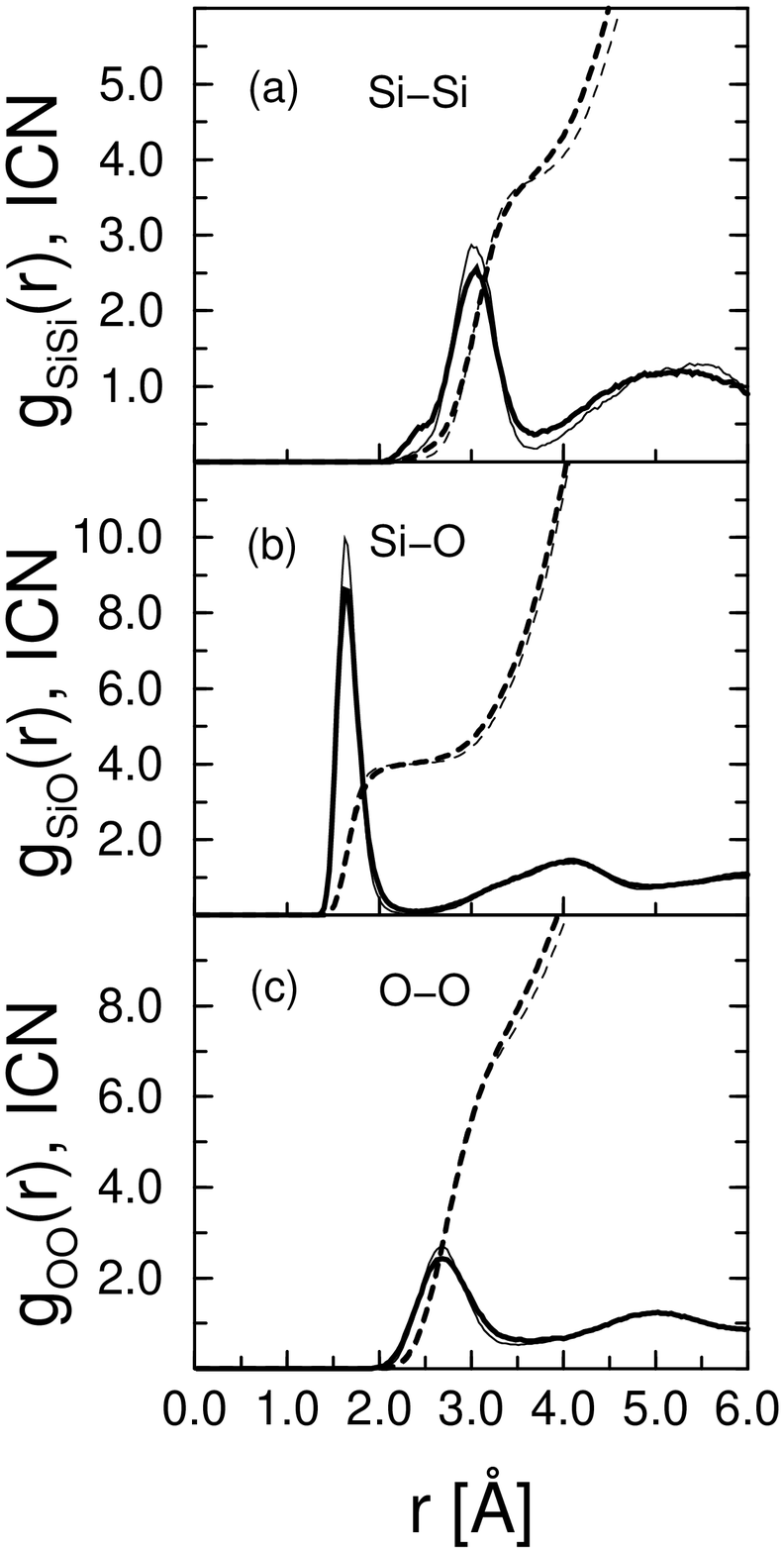}   
 \includegraphics[width=5cm]{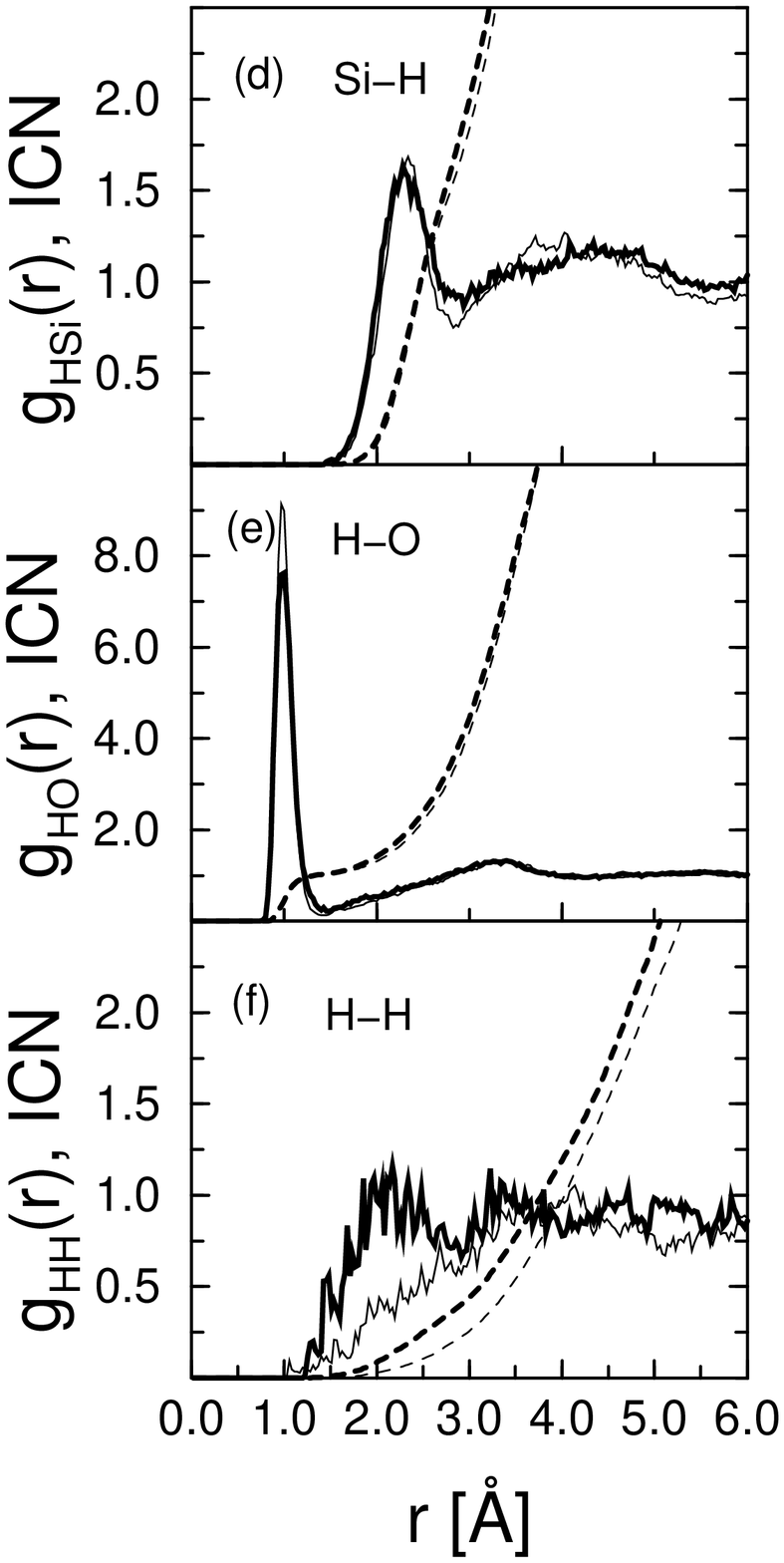} } 
\caption{\label{fig:gdr3530} Radial distribution function (solid lines) and integrated 
coordination numbers (dashed lines) of 
Si-Si (a), Si-O (b), O-O (c), Si-H (d), H-O (e), and H-H (f) for the SiO$_2$-H$_2$O liquids at 3500 K (bold lines) and 3000 K (thin lines).}
\end{figure}

In this subsection we analyze the short range correlations of the liquid in terms 
of the radial distribution functions (RDF)
\begin{equation}
g_{\alpha\beta}(r)=\frac{D_{\alpha\beta}(r)}{4\pi r^2\:\rho_{\alpha}A_{\beta}\:dr}~~~\mbox{with}~~~~\rho_{\alpha}=\frac{N_{\alpha}}{V}~~~~~\mbox{and}~~~~A_{\beta}=N_{\beta}-\delta_{\alpha\beta} 
\end{equation}
where $D_{\alpha\beta}(r)$ is the number of inter atomic distances between $\alpha$ and $\beta$ atoms found between $r$ and $r+dr$. $N_i$ denotes the number of atoms of each kind $i$ and $V$ is the volume of  the simulation box. 
The corresponding integrated coordination numbers (ICN) are
\begin{equation}
{\rm ICN}_{\alpha\beta}(r)=\frac{1}{A_{\alpha}}\int_0^r D_{\alpha\beta}(r')\:dr' ~~~\mbox{with}~~~~A_{\alpha}=N_{\alpha}-\delta_{\alpha\beta}
\end{equation}
The RDF and ICN are presented in the left panel of
Fig. \protect\ref{fig:gdr3530} for the network forming atoms silicon and oxygen ($\alpha,\beta=$Si,O) 
and in the right panel for the pairs involving H.

From this figure, one recognizes that the distributions become broader as the temperature is increased. 
The RDFs involving the matrix atoms show a first peak followed by a well defined first minimum 
which becomes more pronounced if the temperature is decreased from 3500 K to 3000 K.
In particular, the Si-O RDFs present, after the first peak, a very well defined  
minimum  at 2.37 $\pm$ 0.05 \AA\ for 3500 K and at 2.35 $\pm$ 0.05 \AA\ for 3000 K. 
From the position of the Si-O first peak, we can deduce that the most probable Si-O distance
 is around 1.65 $\pm$ 0.02 \AA\ for 3500 K and around 1.63 $\pm$ 0.02 \AA\ for 3000 K. \\
The ICN for Si-O exhibits a plateau at a value of 4 which indicates that every
silicon atom has on average four oxygen neighbors and hence that the principal units - the SiO$_4$ tetrahedra - 
are preserved also in the presence of water. Indeed only a small percentage of threefold and fivefold coordinated Si atoms
are found in the liquids (7 \% fivefold corrdinated and 4 \% threefold coordinated at 3500 K and 2 \% fivefold coordinated and 1 \% threefold coordinated at 3000 K). 
On the other hand,  the ICN for Si-Si shows an inflection point at around 3.6, a value which is smaller than 
the one for a perfect tetrahedral network, 4.0, indicating that the tetrahedral network is partially broken. \\
Comparing the matrix distributions to those of the pure silica melt, we note that the addition of water does not alter signicantly  the shapes of the Si-O, O-O and Si-Si RDFs presented in \cite{Be01}, as it was already the case in the sodium silicate melt \cite{Is02} upon the  addition of sodium.  \\

The RDFs for H-Si and H-O, right panel of Fig. \ref{fig:gdr3530}, 
show somewhat  better  defined inter atomic distances as the temperature is decreased, as it was the case for 
the Si-O, Si-Si and O-O RDFs. On the other hand the H-H distribution seems to deviate from this behavior. 
Concerning the H-Si distribution, the first maximum is found around 2.3 \AA . 
Since this value is much larger than the most probable Si-O distances (1.65 $\pm$ 0.02 \AA\ and 1.63 $\pm$ 0.02 \AA\ ), the presence of stable molecular Si-H units is excluded. 
The height of the first peak in Si-H and the absence of a well-defined minimum 
reflect the presence of a broad distribution of these distances in the liquid. 
In contrast, the well-defined first peak in the H-O distribution functions, located at 0.99 $\pm$ 0.01 \AA , followed by a
 well defined minimum at 1.48 $\pm$ 0.03 \AA\ at 3500 K and at 1.43 $\pm$ 0.03 \AA\ at 3000 K, reveals the O-H bond as 
the dominant stable configuration for the hydrogen atoms. 

Because of a lack of a well-defined first peak in the RDF for H-H, the existence of stable H$_2$ 
molecules can be excluded too. The striking point in the H-H distribution is the difference between 
the RDF at 3000 K and 3500 K between  1.5 \AA\  and  3.0 \AA. Whereas at 3000 K 
there are almost no H-H pairs with a distance around 2.0 \AA\, we find a weak peak at this distance at 3500 K.
This distance is very close to the H-H distance in a free water molecule and hence we have evidence that at this temperature water molecules do exist. 
Indeed, in a more detailed analysis of the coordination numbers as a function of time, 
we found that oxygen atoms with two nearest neighbor 
hydrogens were stable over times of the order of 500 fs. The absence of this phenomenon at 3000 K 
is probably due to  the considerably lower displacement of the hydrogen 
atoms (see Fig. \protect\ref{fig:msd}) at this temperature which, because of a too small trajectory length, 
prevents to find two hydrogen atoms sufficiently close to each other to form a water molecule. \\

We recall at this point that the configuration of the valence shell is the same  for the hydrogen and the sodium atoms
(although it is well known that the physical and chemical 
properties of compounds of equivalent stoichiometry  involving these two atom types (such as H$_2$O and Na$_2$O) 
are rather different). This behavior motivates a comparison of the results of this study to those of the liquid sodium silicate \cite{Is02}. 
\begin{figure}[ht]
\centerline{\includegraphics[width=5cm]{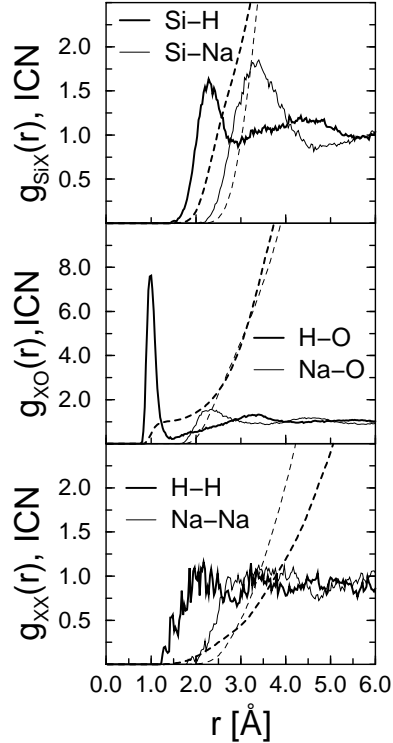}    }
\caption{\label{fig:gdrHNa} Si-X, X-O, X-X (X=H,Na) radial  distribution functions (solid lines) and integrated coordination numbers (dashed lines) for the hydrous silica (bold lines) and the NS4 melt from Ref. \protect\cite{Is02} (thin lines) at 3500 K.}
\end{figure}
In Fig.~\protect\ref{fig:gdrHNa}, we present the comparison of the above discussed hydrogen-containing
 RDFs with the corresponding Na-containing RDFs in the sodium tetrasilicate melt at 3500 K 
from Ref. \cite{Is02}. 
Two main properties seem to govern the character of the distributions: The  size of the atoms 
and the ability of the two atom types to form either covalent or ionic bonds. In particular, 
the RDF for Si-X and X-X (X = Na, H) shown in Fig. \protect\ref{fig:gdrHNa} look quantitatively quite similar 
except that the peaks are shifted to larger distances in the case of Na. 
This behavior can be easily related to the atom size. In contrast to this, 
the O-X distribution function for X=H has a very different shape from the one for X=Na.  
As described above, for the O-H correlation, a sharp peak at an O-H distance of 0.99 \AA\ $\pm$ 0.01 is followed by a 
well-defined minimum. In contrast to this,  the first peak of the O-Na RDFs is much broader with a
maximum at 2.32 $\pm$ 0.05 \AA\ followed by a shallow
minimum around 3.6~\AA . These latter differences  are  the signature of 
the strong covalency of the  OH bond and the ionicity of the Na-O bond.  \\

\begin{figure}[ht]
\centerline{\includegraphics[width=5cm,angle=-90]{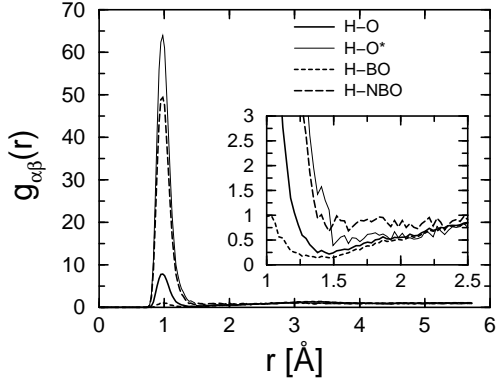}}    
\caption{\label{fig:specOH} Pair distribution functions at 3500 K for H-O (bold solid line), H-BO (dashed line), 
H-NBO (long dashed line) and H-O$^*$ (thin solid line). Inset: Zoom of the minima of the radial distribution functions after
the first peaks.}
\end{figure}

To conclude the section on the radial  distribution functions, we want to investigate how the different oxygen types defined above contribute to the RDF. In order to count the number of Si and H neighbors 
of a given oxygen atom, we used cutoff distances extracted from the first minima of the Si-O and H-O 
radial distribution functions, respectively (cutoff for SiO = 2.37 \AA\ and 2.35 \AA\ and cutoff for HO=1.48 \AA\ and 1.42 \AA\ for the higher and the lower temperature, respectively).
Figure \protect\ref{fig:specOH} shows the H-O radial distribution
functions for the different oxygen types (O$^*$, NBO, BO) compared to the total H-O RDF at 3500 K. 
Results obtained for 3000 K are very similar and are therefore not shown here.\\

We note that the RDFs for all the different oxygen types show a pronounced peak around 1.0 \AA\  but 
that the height of this first peak depends strongly on the considered oxygen type. The H-O$^*$ RDF
exhibits a first peak of height 65 which is much larger than the one for the total H-O which is around 8. 
In contrast to this, the RDF for H-BO has a first peak of height 1. The most striking point 
in Fig. \protect\ref{fig:specOH} is that the peak for H-O$^*$ is significantly higher than 
the one for H-NBO. Since all O$^*$ atoms are also NBO atoms 
(remember that by definition the O$^*$ atoms have a hydrogen atom as second neighbor whereas for 
the NBO atoms this is not necessarily the case), the presence of a large number of Si-O dangling 
bonds becomes now evident. We will discuss the existence and the temperature dependence of these dangling bonds below. 
From the inset of Fig. \protect\ref{fig:specOH}, it can be concluded that the H-NBO contribution is dominant 
for distances between 1.5 \AA\ - 2.5 \AA . This means that the NBO atoms are also connected to hydrogen 
atoms via the so-called hydrogen bonds the length of which is equal to 2.0 \AA\ in the water dimer. 
Though also the RDF for the other oxygen types show a contribution in this range, it is less important. 

\begin{figure}[ht]
\centerline{\includegraphics[width=6cm,angle=-90]{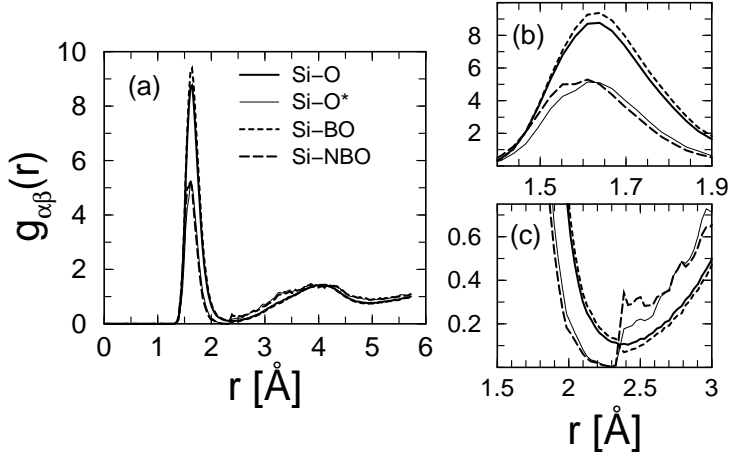} }   
\caption{\label{fig:specOSi} Pari distribution functions at 3500 K for Si-O (bold solid line), Si-BO (dotted line), Si-NBO (dashed line), and Si-O$^*$ (thin solid line). (b) and (c) are magnifications of the first peaks and first
minima, respectively.}
\end{figure}

In Fig. \protect\ref{fig:specOSi}, we present the contribution of the different oxygen types 
to the Si-O radial distribution function.
The height of the first peak of the Si-BO RDF is around  9.3 whereas the peak height of the total Si-O RDF
is close to 8.7 which indicates that the correlation of the bridging oxygen with silicon atoms is 
stronger than the total Si-O correlation (see Fig. \protect\ref{fig:specOSi}b).  
The heights of the first peak in the Si-O$^*$ and Si-NBO RDFs are very close to each other at a value of $\approx$ 5.1. 
As shown in Fig. \protect\ref{fig:specOSi}b, a slight shift
of the peak positions can be observed: The Si-NBO and Si-O$^*$ peak positions are around 1.60 \AA\ and
 1.62 \AA , respectively, whereas the Si-BO peak position is located at 1.63 \AA . This result implies that
the tetrahedra having one or more NBO atoms are distorted, as it has already been observed in other 
silicate systems (see Ref. \cite{Is01} and references therein). \\ 
Figure \protect\ref{fig:specOSi}c shows the presence of a considerable jump in the RDFs for Si-O$^*$ and Si-NBO at the Si-O cutoff. Between 2.37 \AA\ (the Si-O cutoff) and  4.0 \AA , the Si-O$^*$ and Si-NBO RDFs are larger than the total one
(Fig. \protect\ref{fig:specOSi}c). 
This jump and the dominance of the Si-O$^*$ and Si-NBO can be associated to oxygens of the  O$^*$  type or to 
NBOs that are connected to a second silicon atom by a weak Si-O bond. As soon as the distance to the second 
silicon atom becomes smaller than the cutoff (i.e. an O3H or BO is formed) the oxygen is considered as BO 
and therefore the Si-O$^*$ or Si-NBO distribution drops abruptly to zero at the cutoff distance. Thus this behavior 
gives insight into the formation and decay processes of the O3H clusters and the Si-O dangling bonds, 
that will turn out to be one of the essential transition states for hydrogen diffusion. \\

\subsection{Angular Distributions}
\begin{figure}[hit]
\centerline{\includegraphics[width=5cm]{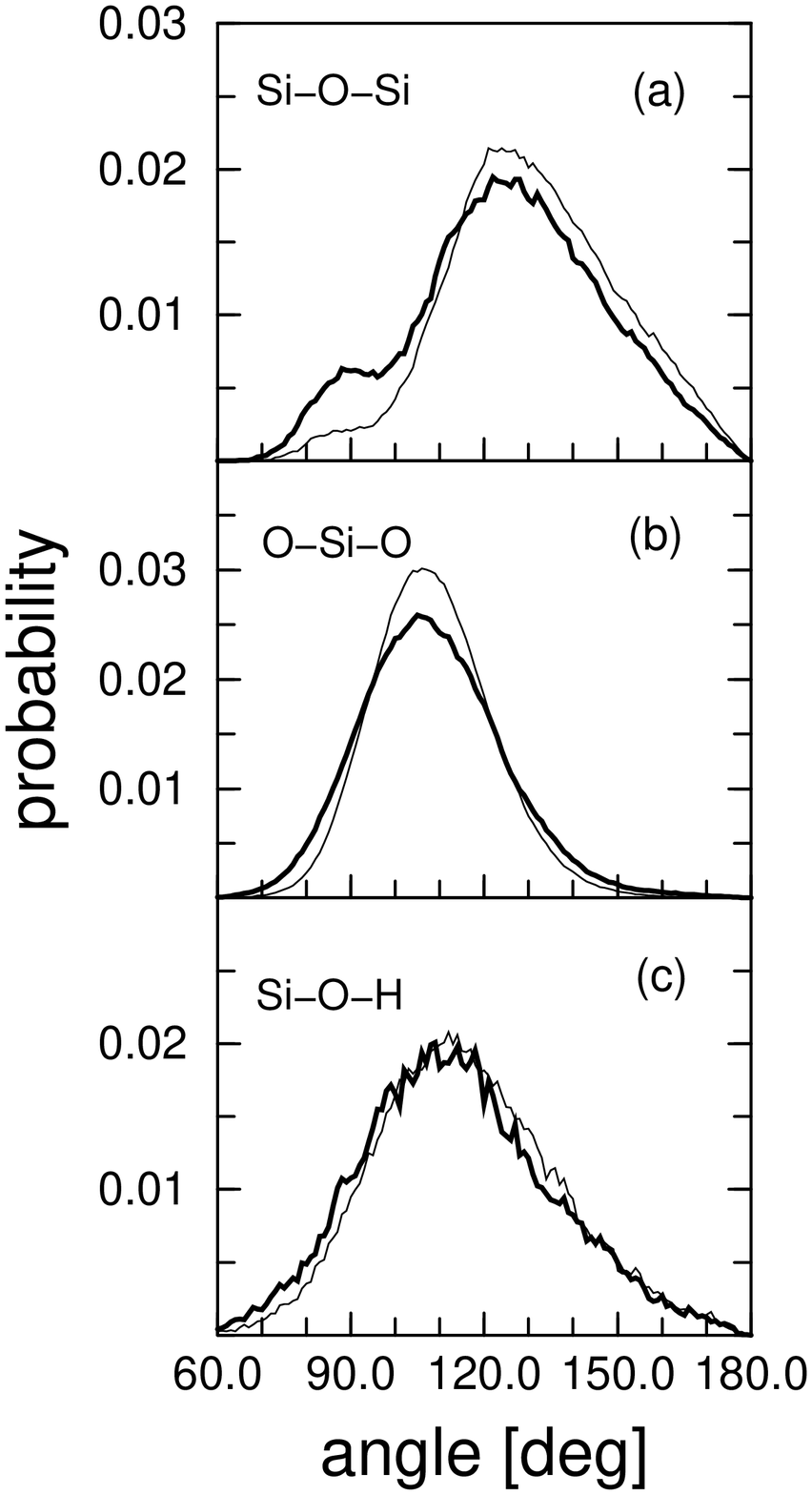} 
\includegraphics[width=5cm]{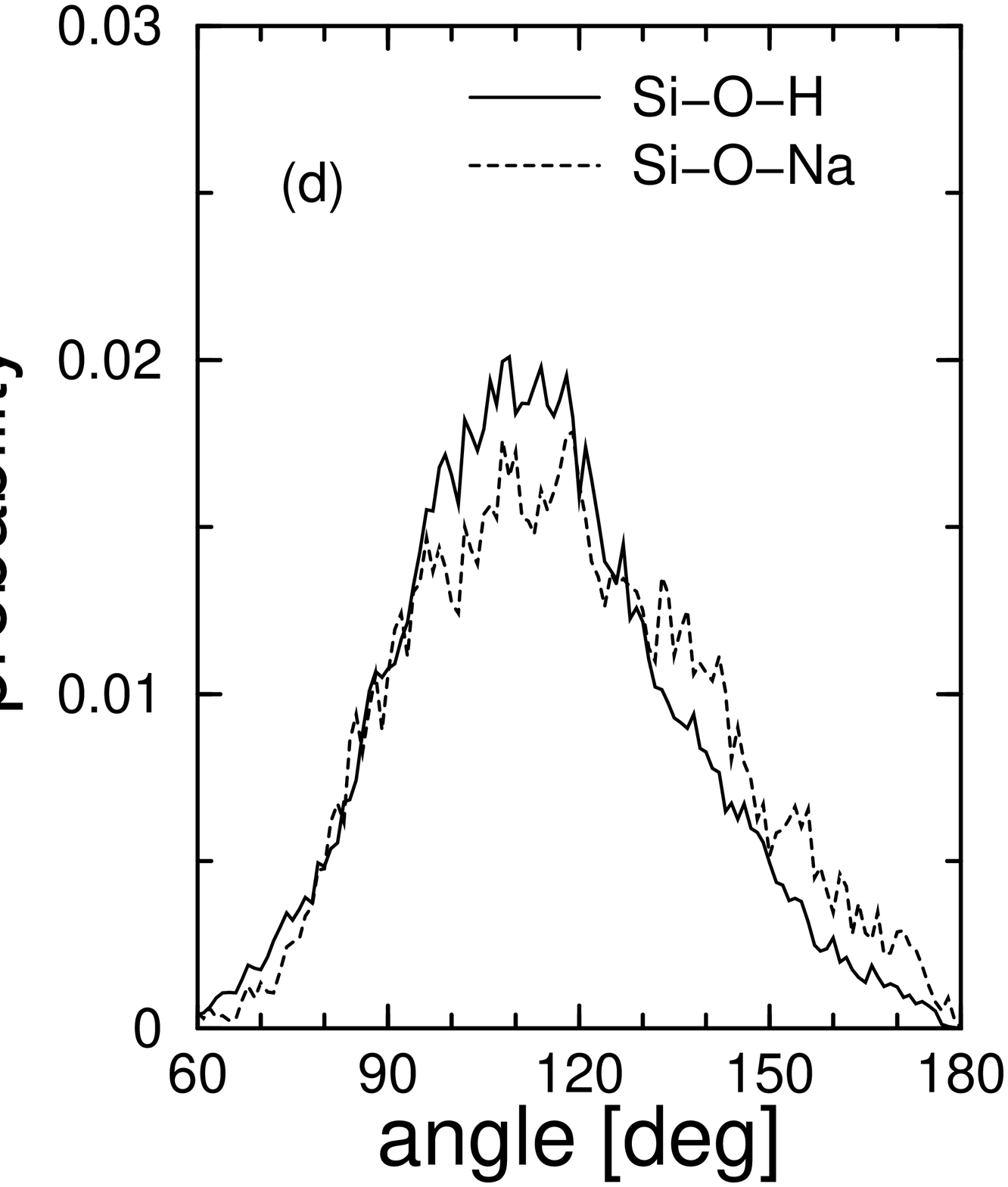}    }
\caption{\label{fig:ang3530} (a)-(c) Angular distributions of the hydrous silica 
sample at 3500 K (bold lines) and 3000 K (thin lines); (d) comparison to NS4 at 3500K (dashed line).}
\end{figure}

Figure \protect\ref{fig:ang3530} presents the different angular distributions in the SiO$_2$-H$_2$O liquids 
at 3500 K and 3000 K and the comparison with NS4 at 3500 K. We remark that the angular 
distributions for the network formers (Si-O-Si) and (O-Si-O) are close to the ones of pure silica at 3500 K \cite{Be01}. 
In agreement with the behavior of the RDFs (Sec. \protect\ref{sec:RDF}), the angular distributions at the lower temperature 
are sharper and more peaked. The Si-O-Si distribution shows an additional hump at $90^{\circ}$ that is also present in pure
silica at 3500 K but much less pronounced. This hump is due to the two membered rings which are formed by two SiO$_4$ units
connected by an edge and having two common oxygen atoms. 
 As will be seen in Sec. \protect\ref{sec:FND}, these units are related to intermediate states 
that are relevant for the diffusion process. From the Si-O-Si distribution in Fig.
\protect\ref{fig:ang3530}, we note that the number of such rings increases with increasing temperature.
The O-Si-O angular distributions present a well-defined maximum at $\approx$ 110$^{\circ}$ which
corresponds to a typical intra-tetrahedra angular distribution (ideal tetrahedra angle $\approx$ 109$^{\circ}$). 
Finally, we can recognize that the temperature dependence of the Si-O-H distribution seems 
to be significantly weaker than the one of the Si-O-Si and O-Si-O distributions. \\
The Si-O-X (X= H, Na) angular distributions at 3500 K (Fig. \protect\ref{fig:ang3530}d) 
present very similar shapes with a slight shift to larger angles for the sodium silicate liquid. 
In both cases we see a broad maximum at angles between 108$^{\circ}$ and 115$^{\circ}$.  
%Note that this range does not contain the H-O-H angle in a water molecule of 104$^{\circ}$.
A maximum in this range shows that, in both cases,
 the chemistry of a twofold coordinated oxygen with a lone pair  is realized.
 The shift of the mean angle to a higher value for the sodium 
silicate is certainly an effect of the larger size of the sodium atom since for larger atoms 
the repulsion of the electron shells of the two oxygen-ligand atoms becomes important.  \\

\subsection{Structure Factors}

\begin{figure}[ht]
\centerline{\includegraphics[width=9cm]{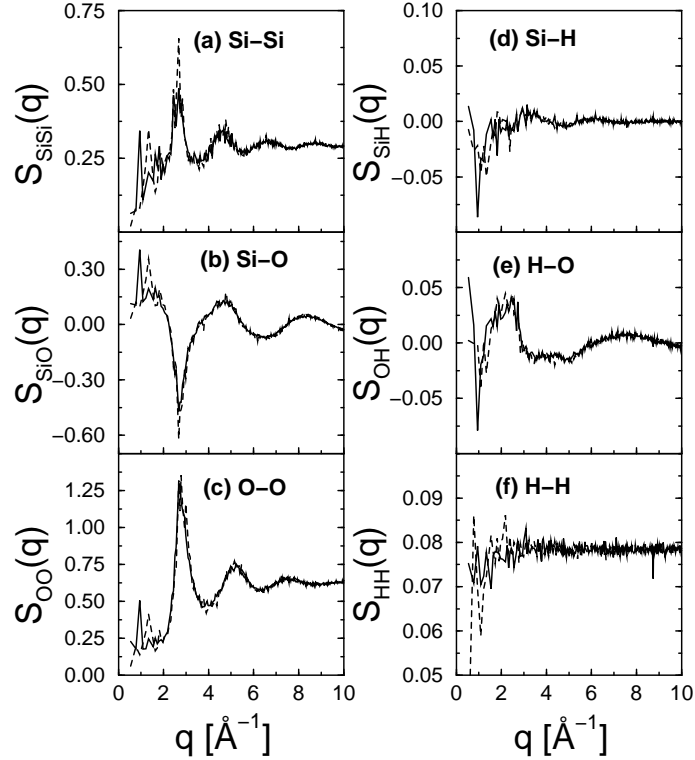}    }
\caption{\label{fig:Spartq} Partial structure factors of the hydrous silica liquids
at 3000 K (dashed line) and at 3500 K (bold line).}
\end{figure}

We calculated the neutron scattering structure factor according to
\begin{equation}
S_n(q)=\frac{1}{N_{\rm Si}b_{\rm Si}^2+N_{\rm O}b_{\rm O}^2+N_{\rm H}b_{\rm H}^2}\sum_{k,l}^{N}b_kb_l\langle 
\exp[i{\bf q}\cdot({\bf r}_k - {\bf r}_l)]\rangle
\end{equation}
where $b_k$ ($k$=Si,O,H) are the coherent neutron scattering lengths and $\langle \cdot \rangle$ 
is the thermal average. The scattering lengths were taken from Ref. \cite{Ni95} where 
$b_{\rm Si}=0.41491\times 10^{-14}$ m, $b_{\rm O}=0.5803\times 10^{-14}$ m and $b_{\rm H}=-0.374 \times 10^{-14}$ m are reported. A very important feature for the experimental verification of the simulation is the replacement of hydrogen by its isotope deuterium. 
Since the coherent scattering length of deuterium ($d_{\rm D}=0.671\times 10^{-14}$ m) is quite different from that of hydrogen,
 the contribution of the hydrogen atoms to the total structure factor can be revealed.   
\begin{figure}[ht]
\centerline{\includegraphics[width=5cm,angle=-90]{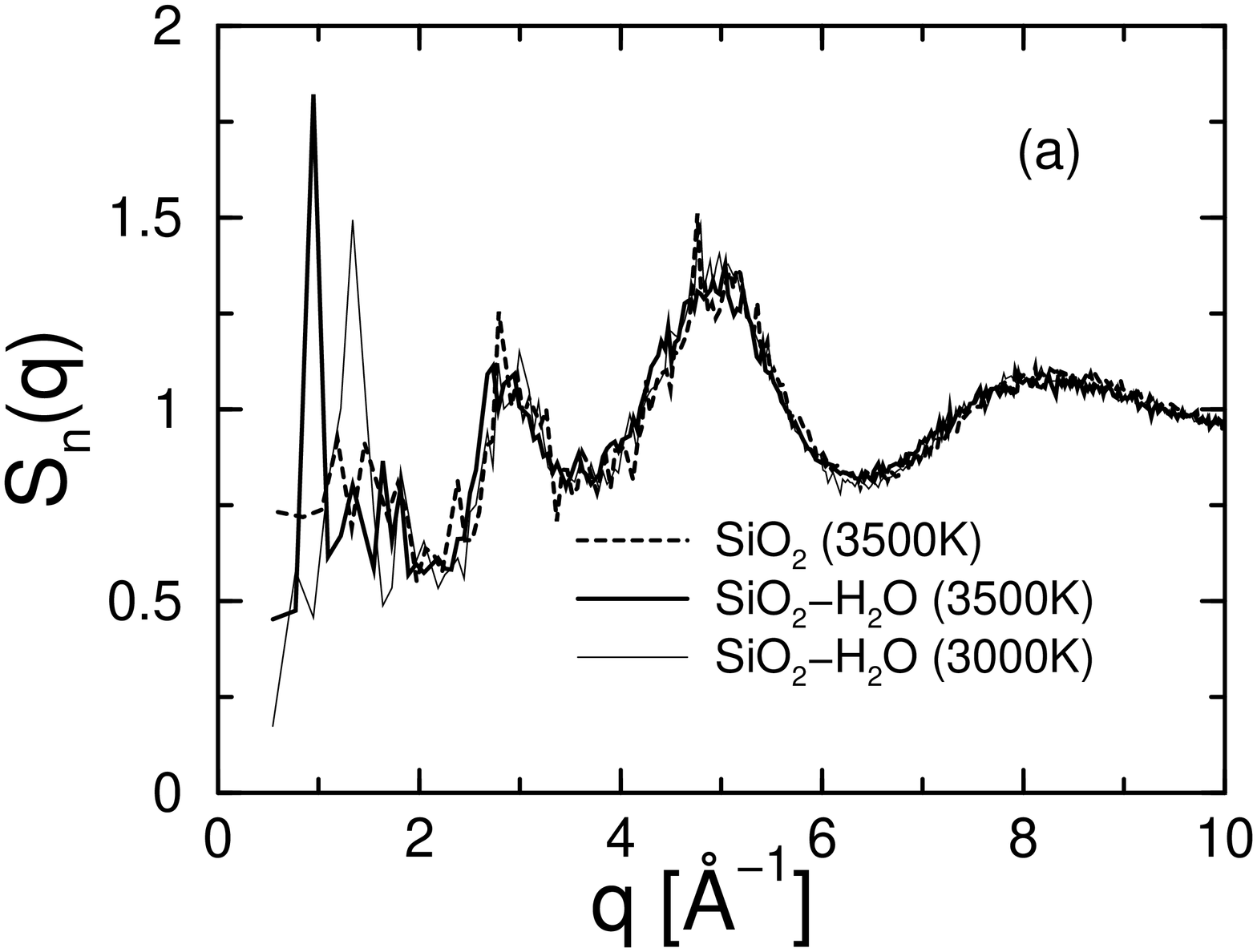} 
\includegraphics[width=5cm,angle=-90]{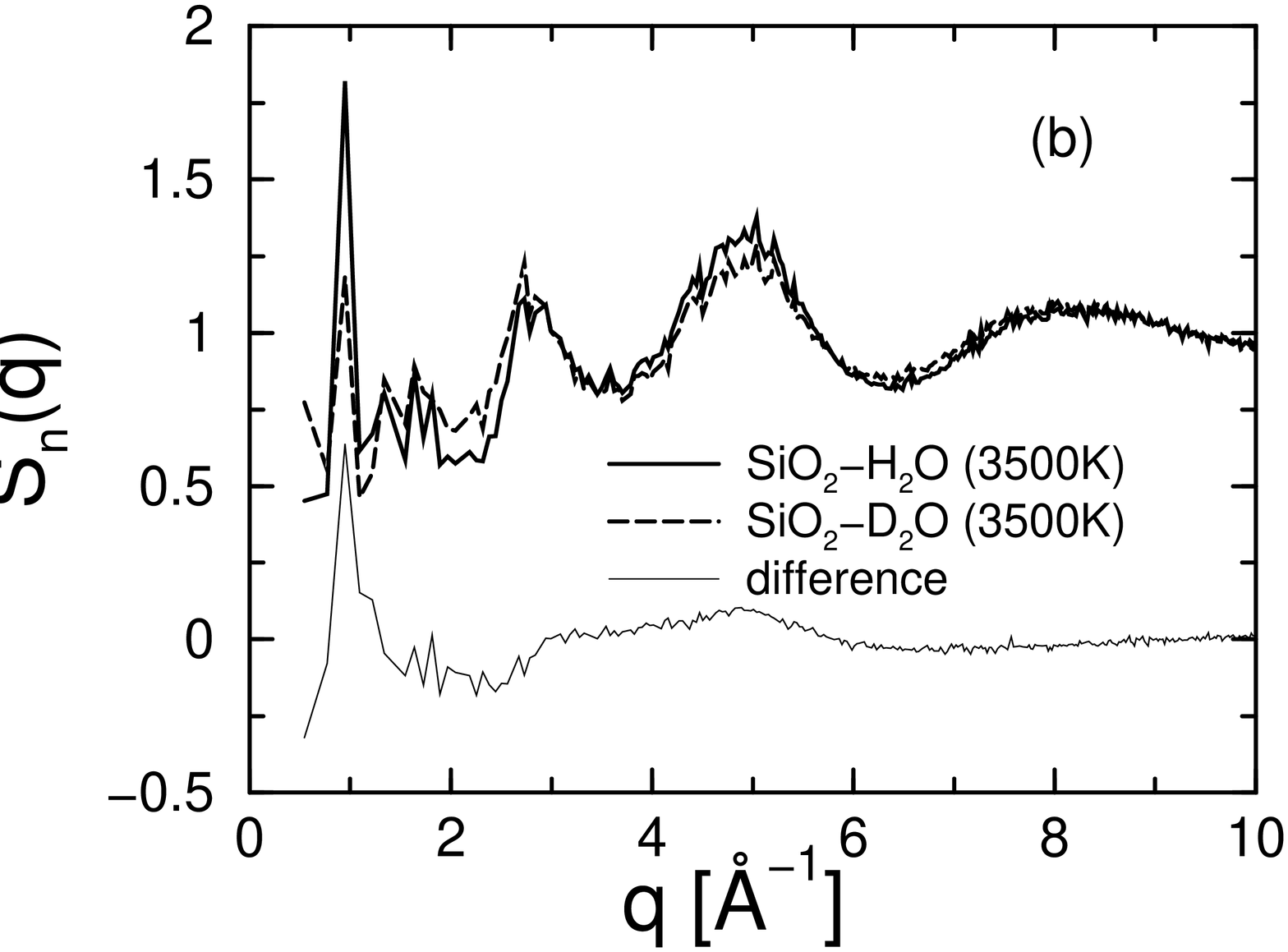}   } 
\caption{\label{fig:Stotq} {(a)} Total neutron scattering structure factor of the hydrous silica liquids
at 3000 K (thin line) and at 3500 K (bold line) and of a pure silica liquid at 3500 K (dashed line); {(b)} Total neutron scattering structure factor for the hydrous silica liquid at 3500 K (bold solid line), the deuterated liquid (bold dashed line) 
and the difference (thin solid line).}
\end{figure}
Figure \protect\ref{fig:Stotq}a  shows the neutron scattering structure factor $S_n(q)$ for both simulated 
temperatures in comparison to the simulated dry silica liquid at 3500 K \cite{Be01}. 
For $q > 2.0$ \AA $^{-1}$, the three curves can be considered to be identical which is in agreement with the fact that
 we did not see any differences in the Si-Si, Si-O and O-O radial distribution functions for silica, the sodium silicate, and the hydrous silica liquids. In contrast to this, the $q\leq 2.0$ \AA\ $^{-1}$ 
region of $S_n(q)$ of the hydrous silica liquid exhibits a prepeak at 0.95 \AA $^{-1}$ at 3500 K and at 1.3 \AA $^{-1}$ 
at 3000 K which is not present in the pure silica melt.  (We mention that the box size  of  the hydrous silica liquid is 11.5 \AA ,  
corresponding to a minimum value of $q$ of 0.55 \AA $^{-1}$  and thus the prepeak found at 0.95 \AA $^{-1}$ 
can not be attributed to a size effect. Note also that the number of silicon and oxygen atoms was almost 
equivalent in the simulation of pure silica and hence the statistical accuracy is as well comparable.) \\
For the sodium silicate it turned out that the long range correlations 
($q\leq 2.0$ \AA $^{-1}$) contain  highly important information on the diffusion mechanism as it was shown 
recently \cite{Ho01,Ho02,Me02}. For the sodium silicate it turned out that the long range
correlations  ($q\leq 2.0$ \AA $^{-1}$) contain important information on
the diffusion mechanism as it was shown recently \cite{Ho01,Ho02,Me02}. It
was found that in the liquid a network of channels is formed which
enables the sodium atoms to move rapidly through the SiO-matrix. This
channel structure is thus able to explain the relatively high diffusion
constant of sodium atoms in sodium silicate melts. The characteristic
distance between these channels is around 6 \AA, i.e. two tetrahedra
diameters, and gives rise to a prepeak in the structure factor at about
1 \AA $^{-1}$, a structural feature which has indeed be found in recent
neutron scattering experiments \cite{Me02}. \\
Fig. \protect\ref{fig:Stotq}b shows the structure factor at 3500 K for the hydrated silica and deuterated silica liquids
as well as their difference. We note remarkable differences of the two signals at 0.95 \AA $^{-1}$, around 2.0 \AA $^{-1}$ and around 5.0 \AA $^{-1}$. 
Since the most important contributions to the difference are located in the $q$-vector region of the prepeak 
(at 0.95 \AA $^{-1}$), the prepeak can be directly attributed to the presence of the hydrogen atoms. 
For a deeper understanding of the prepeak origin in the hydrous melts, we analyzed the partial 
structure factors presented in Fig. \protect\ref{fig:Spartq}.  
 
The partial structure factors $S_{\alpha\beta}(q)$ are related to the total $S_n(q)$ presented 
in Fig. \protect\ref{fig:Stotq} as follows:
\begin{equation}
S_n(q)=\frac{1}{\sum_{\alpha}N_{\alpha}b_{\alpha}^2}{\sum_{\alpha\beta}^{N}b_{\alpha}b_{\beta}} S_{\alpha\beta}(q)
\end{equation}
where $S_{\alpha\beta}(q)$ are given by 
\begin{equation}
S_{\alpha\beta}(q)=\frac{f_{\alpha\beta}}{N}\sum_{k}^{N_{\alpha}}\sum_{l}^{N_{\beta}} \langle \exp[i{\bf q}\cdot({\bf r}_k - {\bf r}_l)]\rangle
\end{equation}
and $f_{\alpha\beta}$ is equal to 0.5 for $\alpha\neq\beta$ and equal to 1.0 for $\alpha=\beta$. 
The partial structure factors in Fig. \protect\ref{fig:Spartq} can be separated in a group describing 
the silica matrix distributions (Si-Si, Si-O, O-O) and in a group involving distributions with H 
(Si-H, O-H, H-H) as we did for the RDFs (see Sec. \protect\ref{sec:RDF}). For the matrix part, 
we find for $q > 2.0$ \AA\ $^{-1}$ a nearly perfect agreement between the structure 
factors presented in Fig. \protect\ref{fig:Spartq} 
and the ones extracted from the NS4 liquid simulation from Ref. \cite{Is02}. 
The only difference is  the prepeak at 0.95 \AA $^{-1}$ for 3500 K and at 1.3 \AA $^{-1}$ for 3000 K. 
We recognize therefore a modification of the matrix at a length scale between 4.8 \AA\ and 6.6 \AA\ which 
was not visible in the RDF representation. Since the prepeak is present not only in the partial structure factors 
involving hydrogen, but also in the O-O and Si-Si distributions, we note that the network modification does concern not only
the hydrogen atoms. This behavior can be understood by taking into account the strong and well defined bonding 
of the hydrogen atoms to the silica network (see the  O-H radial distribution function in Fig. \protect\ref{fig:gdr3530}). \\

\subsection{Distribution of Nearest Neighbors}
\label{sec:FND}

\begin{figure}[ht]
\begin{center}
\includegraphics[width=13cm]{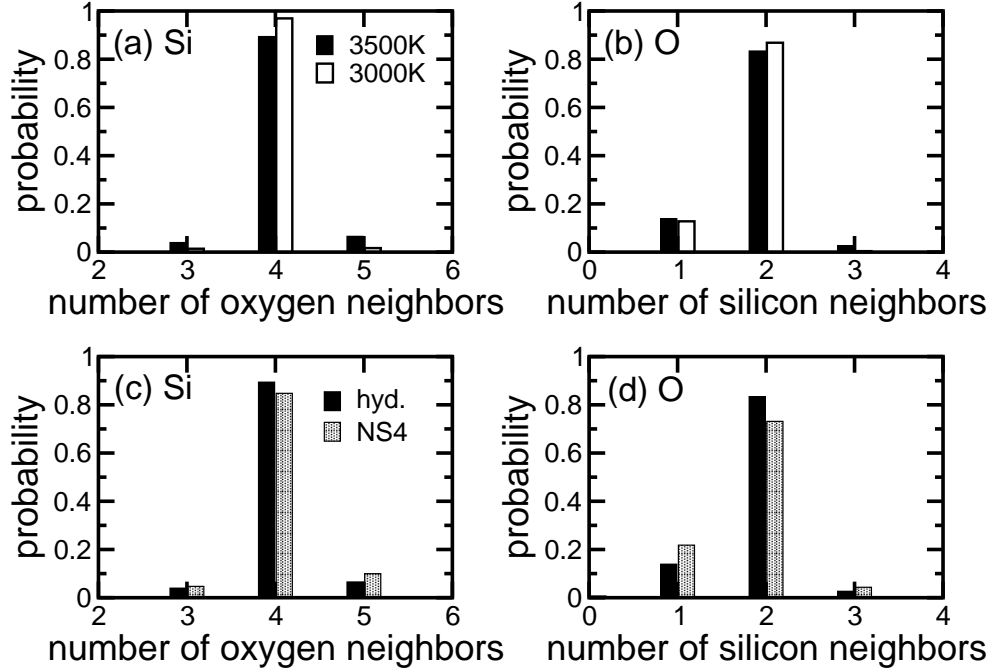}    
\end{center}
\caption{\label{fig:coordtemp} Si-O (a) and O-Si (b) coordination distributions 
for the hydrous silica liquids at 3500 K (black bar) and 3000 K (white bar) and  Si-O (c) and O-Si (d) for the sodium tetra-silicate at 3500 K (grey bar). The black bars in (c) and (d) are identical to the ones in (a) and (b) and represent the hydrous liquid at 3500 K. }
\end{figure}

In this section we discuss the nearest neighbor coordination numbers. 
The nearest neighbors of an atom were again determined as the atoms that are located 
within a sphere with a radius that is given by the positions of the first minima in the corresponding radial distribution function. 
Whereas the average coordination number has already been extracted from the RDFs themselves 
(Fig. \protect\ref{fig:gdr3530}), 
we present now the distributions. The discussion is limited to the coordinations 
of the network forming atoms. The errors for the results presented in this section are related to the number of 
independent configurations in the liquid. We assume a configuration to be independent of a previous one 
if they are separated by the time it takes for the atoms to show a diffusive motion. 
From Fig. \protect\ref{fig:msd} we thus recognize that for T$=$3000 K and T$=$3500 K we have two 
and three independent configurations, repectively.
Thus the relative errors are $1/\sqrt{2}=0.71$ at 3000 K and 
$1/\sqrt{3}=0.58$ at 3500 K and are then divided by the square root to the considered average
quantities. The resulting absolute errors will be given in the text. \\
As for the other quantities described above, we begin with the temperature dependence
of the distributions in the hydrous liquid and extend the study of the comparison of these distributions 
with the ones found in the sodium silicate liquid. \\
Figures \protect\ref{fig:coordtemp}a and b present the Si-O and O-Si distributions for the two temperatures 
of the hydrous silica liquid. Both coordination probabilities confirm the picture drawn from the analysis of the previous 
quantities: The silica network is still present in the hydrous silica liquid since we find a maximum of the Si-O 
distribution at a coordination of four and a maximum of the O-Si distribution at a coordination of two.\\

In agreement with the temperature dependence of the RDFs and of the angle distributions, 
the coordination distributions become broader as the temperature is increased. 
For the higher temperature, the Si-O distribution shows significant contributions at coordination numbers of 
three and five and the O-Si distribution shows contributions at a coordination of one.\\
In addition, the O-Si coordination reveals the speciation of the water molecules, since we do not find a 
significant contribution at a coordination number of zero. Therefore the existence of free water molecules can be excluded 
(a free water molecule has zero silicon neighbor). Note that in reality the numerical value is non-zero 
(but too small to be visible in the figure) because of the presence of {\it transient}
 water molecules in the liquid at 3500 K. 
In contrast, the contribution at a coordination of one is roughly $\frac{1}{8}$ at both temperatures, corresponding to the 
fraction of the non-bridging oxygen in the system that compensate the charges of the eight hydrogen atoms 
in the form of O-H groups. \\    
Since the dissolution product of water in silica is almost exclusively made of Si-OH units, 
the dissolution mechanism is similar to that of disodium oxide in the NS4 liquid. 
To compare these two mechanisms, we present the Si-O and O-Si coordination numbers of the hydrous silica liquid
and of the sodium tetrasilicate liquid at 3500 K in Figs. \protect\ref{fig:coordtemp}c and d.
When comparing these two systems, the concentration of sodium atoms of 13.3 mol \% in the NS4 melt  and 
the concentration of hydrogen atoms of 7.8 mol \% in the hydrous silica melt should be taken into account. 
The Si-O coordination distribution has basically the same shape for both systems with a clear maximum at 
a coordination number of 4 even though the distribution for NS4 is, however, somewhat broader. 
The O-Si coordination probability presents stronger differences. Both distributions have their maximum at 2 
(indicating that in both cases the silica network is still present) but the absolute values are quite different. 
In both cases the O-Si probability for having one silicon neighbor deviates significantly from zero. This indicates the 
formation of O$^*$ in the hydrous melt and equivalent oxygen types in the NS4 melt (hereafter denoted by NaO$^*$). 
If the dissolution product was exclusively given by O$^*$ and NaO$^*$, the coordination probability would 
correspond to the N$_{\rm H}$/N$_{\rm O}$ ratio or to the N$_{\rm Na}$/N$_{\rm O}$ ratio, respectively,
where N$_{\rm H}$, N$_{\rm O}$ and N$_{\rm Na}$ are the number of hydrogen, oxygen or sodium atoms in the different systems.
These ratio are equal to $\frac{12}{54}=0.222$ for the sodium silicate liquid and to 
$\frac{8}{64}=0.125$ in the hydrous silica one. Figure \ref{fig:coordtemp}d shows a contribution  
of 0.139 $\pm$ 0.027 
for the hydrous silica sample and of 0.218 $\pm$ 0.037 for the NS4 sample which is in agreement with the
simple theoretical prediction.
In the hydrous case, we note again the important presence of Si-O dangling bonds at 3500 K compared to 
the NS4 liquid at the same temperature.
 
A quantity closely related to the O-Si and Si-O coordinations is the distribution of the $Q^n$ species, 
where $n$ denotes the number of bridging oxygens attached to a silicon atom. For the simulated systems, 
we note that for a perfect dissolution of the water into OH groups, one expects a probability for $Q^3$ 
equal to $\frac{8}{30}=0.267$, the ratio between the number of  hydrogen and silicon atoms,  
if all OH groups are attached to different silicon atoms. 
\begin{figure}[h]
\centerline{\includegraphics[width=5cm,angle=-90]{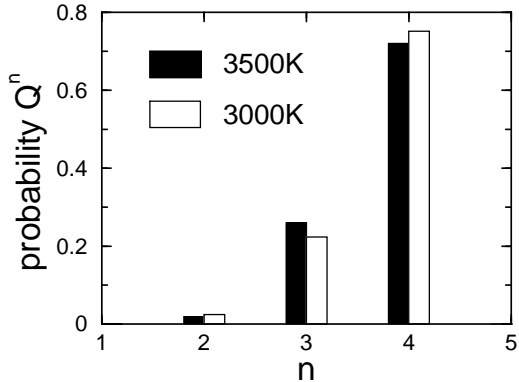}    }
\caption{\label{fig:Qn} Probability to have a $Q^n$ speciation for the hydrous sample at 3500 K (black bars) and 3000 K (white bars).}
\end{figure} 
Indeed Fig. \protect\ref{fig:Qn} exhibits a probability of $Q^3$ sites of 0.229 $\pm$ 0.062 at 
3000 K and of 0.269 $\pm$ 0.055 at 3500 K, respectively.
 The contributions at $n=2$ of 0.029 $\pm$ 0.022 at 3000 K and of 0.023 $\pm$ 0.016 at 3500 K are 
relatively small. In particular, these
contributions are smaller than $\frac{8}{30}\frac{7}{30}=0.062$, the probability that two H among the 
eight ones of the system are found 
on the same Si tetrahedra, 
which indicates that the Si-O-H groups tend to avoid each other. 
Due to the absence of contributions at $n=1$ and $n=0$, we exclude the possibility for a clustering of O-H groups
on specific silicon atoms. Consequently a relatively homogeneous distribution of the O-H groups over the silicon atoms 
is observed. Besides,  note  that one $Q^2$ site in the system would give a probability of $\frac{1}{30}=0.033$ which
is larger than the probabilities for $Q^2$ presented in Fig. \protect\ref{fig:Qn}. This indicates that the $Q^2$ 
sites do not exist all along the trajectories. \\
The $Q^n$ species distribution is experimentally accessible with NMR spectroscopy. Farnan {\it et al.} \cite{Fa87}
have determined these quantities for hydrous silica glasses from $^{29}$Si NMR spectroscopy. Their samples contained 
2.5 and 8.7 wt.\% H$_2$O and the measured $Q^3$ probabilities were $0.086\pm 0.008$ and 
$0.235\pm 0.025$, respectively, e.g. values that are significantly lower than the $Q^3$ probabilities found in the present study. This might be due to the fact that in the samples of Ref. \cite{Fa87}  only half or less of the water molecules were dissolved into OH groups. 
For the 2.5 wt.\% H$_2$O sample, the authors found a ratio of OH over H$_2$O of 1/1 and an even lower value 
in the 8.7 wt.\% H$_2$O sample. Since the samples were prepared at only 1550$^{\circ}$C, these difference in the concentration of the $Q^3$ species might be 
due to the difference  between the experimental temperature and the present simulation ones.  
      
\begin{figure}[h]
\centerline{\includegraphics[width=4.5cm,angle=-90]{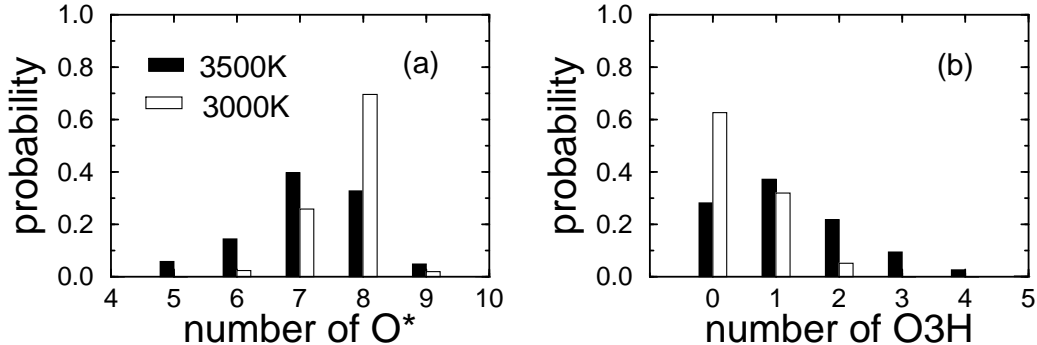}    }
\caption{\label{fig:Odist} Probability to find O$^*$ (a) and O3H cluster (b) in
 the hydrous silica liquid at 3500 K (black bar) and 3000 K (white bar).}
\end{figure}

We also evaluated the probabilities to find oxygen atoms of  different types 
 during the trajectories at 3000 K and 3500 K. 
In particular the appearance of intermediate states like Si-O dangling bonds and O3H triclusters 
as well as water oxygens (wO) can have 
consequences for the hydrogen diffusion in the melt. Figure \protect\ref{fig:Odist} shows the probability to find O$^*$ and O3H cluster units in the trajectories at 3000 K and 3500 K.  

We recall that a dissolution of all hydrogen atoms into O-H groups would give an exact number of eight 
 O$^*$ (the number of H atoms). As it can be seen in Fig. \protect\ref{fig:Odist}a,  
the probability distribution is shifted towards values smaller than eight and that at 3500 K the maximum is not
even at eight. In contrast, the number of O3H cluster increases with  increasing temperature
(Fig. \protect\ref{fig:Odist}b). Since we find that the number of free O-H groups 
and free water molecules is negligible, we conclude that the decrease in concentration of O$^*$ at high temperature is due 
to a direct conversion into O3H clusters. The important re-decay of O3H clusters into O$^*$, which will be
presented in Sec. \protect\ref{subsec:DPH}, seems to confirm this hypothesis.   
From Fig. \protect\ref{fig:coordtemp} and the simulation of pure silica at 3500 K \cite{Be01}, 
we know that three silicon coordinated oxygen atoms are hardly present. The tricluster site formation is 
hence facilitated by the presence of hydrogen atoms. Obviously the O-H groups constitute ``dead-end-pieces'' 
and  a higher angular mobility for an O$^*$ oxygen  than for a BO one can be expected. 
This higher angular mobility enables the approach 
of the O$^*$ to another silicon atom and hence the formation of a tricluster. 
The lower mobility of BO atoms seems to suppress this process in pure silica.
We have also studied the relation of the O3H cluster and the two membered rings 
and found a significant correlation since more than half of the O3H clusters is part of a two membered ring.  
    
\begin{figure}[h]
\centerline{\includegraphics[width=5cm,angle=-90]{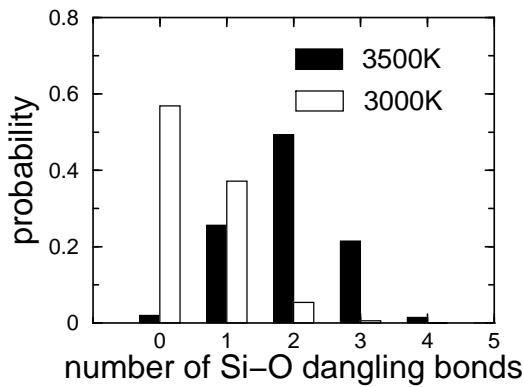}    }
\caption{\label{dang}  Probability distribution of the Si-O dangling bonds for the hydrous silica liquid
at 3500 K (black bars) and 3000 K (white bars).}
\end{figure}

The existence of Si-O dangling bonds has been demonstrated in Fig. \protect\ref{fig:specOH}. 
In Fig. \protect\ref{dang} we present the temperature dependence of the probability to find 
a given number of these oxygen types in the hydrous silica liquid. \\     

Whereas at 3000 K the maximum  of the distribution is at zero,  it shifts to a value of two at 3500 K and 
the probability to find zero dangling bonds at this temperature is small. Hence, the formation of these 
species is an effect of the elevated temperature. Since this structural feature is not found in a pure silica liquid, 
the question of its origin emerges. We relate the existence of these dangling bonds directly to the formation 
of the O3H cluster. Since these tricluster units increase the coordination of
the oxygen atom from two to three, the system attempts to compensate of this stoichiometry 
violation. Since the coordinations of the hydrogen and silicon atoms are rarely violated, 
the appearance of the Si-O dangling bonds (where O has only one neighbor) is certainly associated to
the appearance of the O3H clusters.  \\
Now that the existence of the Si-O dangling bonds is evident, the question of their contribution to an 
eventual fast hydrogen diffusion emerges. 
Do these dangling bonds serve as acceptor and donator states 
for hydrogen ?  From Figs. \protect\ref{fig:specOH} and \protect\ref{fig:specOSi}, we know that
Si-O dangling bonds exist both with a weak bond to a silicon atom and with a weak bond to a 
hydrogen atom. But the analysis of the recombination of these sites 
shows that only about 17 \% of the dangling bonds recombine to a O$^*$ site and 83 \% recombine to a BO site 
(see Sec. \protect\ref{subsec:DPH}).   

\begin{figure}[h]
\centerline{\includegraphics[width=5cm,angle=-90]{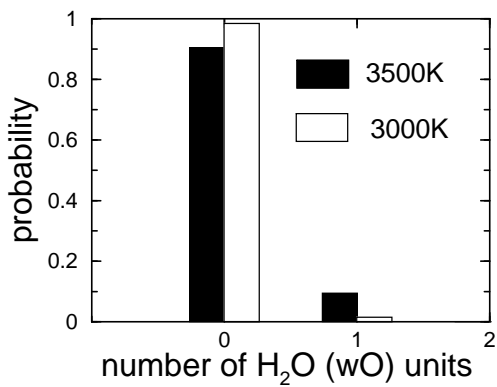}    }
\caption{\label{fig:h2o} Probability for H$_2$O units at 3500K (black bars) and 3000K (white bars).}
\end{figure}

The last oxygen species we want to present here is the water oxygen (wO). As already discussed in the introduction, 
these units are supposed to play a decisive role for the proton transport. We also know from the H-H RDF first
peak at 3500 K (Fig. \protect\ref{fig:gdr3530}) 
that such units exist at this temperature. 
Figure \protect\ref{fig:h2o} shows the probability to find such water units.
Indeed at the higher temperature, one water unit is present with a probability of roughly 10 \%
along the trajectory. At 3000 K this probability is only 2 \%. The corresponding water concentrations 
are 0.30 mol \%  and 0.06 mol \%, respectively. Note that the concentration of 0.30 mol \% at 3500 K should 
give rise to a probability of 0.003 for the zero O-Si coordination in Fig. \protect\ref{fig:coordtemp} 
if the water molecules were free. 
Since the contribution at zero is two orders of magnitude lower, we underline the important point 
that those water units are not free. A detailed analysis reveals an O3 coordination with two hydrogen neighbors and 
one silicon neighbor. As already mentioned above, the found water unit concentrations are far remote 
from experimentally measured water concentrations. Considering again the data of Farnan {\it et al.} \cite{Fa87} 
where a ratio of OH over H$_2$O of 1/1 (50 mol \% H$_2$O) at a total water content of 2.5 wt.\% is found, 
one would expect important contributions around two in Fig. \protect\ref{fig:h2o}.  
Again these differences in the concentration of molecular water probably arise from the difference
between the experimental temperature and the present simulation ones. \\
\\
We conclude the section on the structural properties by a brief summary of the figures that showed evidence of the existence of the different intermediate states and that quantify their probability of occurence in the liquid: 
\begin{table}[h]
\begin{center} 
\begin{tabular}{l|l}
\hline \hline 
{\sf structural unit} &  {\sf relevant figures}  \\ \hline
SiOH groups & 2e (evidence), 11a (quantification)\\
Si-O dangling bonds & 4 (evidence), 5 (evidence), 12 (quantification) \\
O3H triclusters & 5c (evidence), 11b (quantification)\\
water molecules & 2f (evidence), 13 (quantification)\\
\hline \hline
\end{tabular} 
\end{center}

\caption{\label{tab:trasi} Figures that show evidence for the existence of the transition states and that quantify their probability of occurence.}
\end{table}

\subsection{Hydrogen Diffusion Process}
\label{subsec:DPH}

\begin{figure}[h] 
\centerline{\includegraphics[width=13cm]{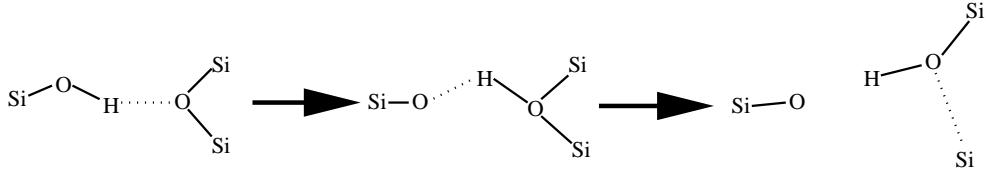}  }
\caption{\label{fig:diffreact1} Typical hydrogen diffusion reaction of process {\bf 1.}. A hydrogen is released from 
an Si-O-H (O$^*$) to a bridging oxygen (BO) forming subsequently another O$^*$. A Si-O dangling bond and 
an unsaturated silicon atom are the resulting products. }
\end{figure}

\begin{figure}[h] 
\centerline{\includegraphics[width=13cm]{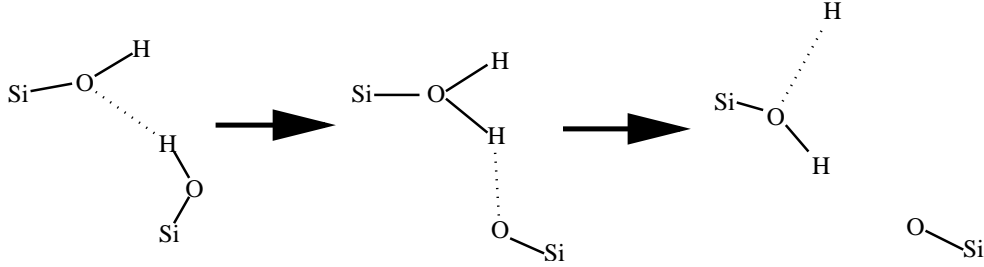}  }
\caption{\label{fig:diffreact2} Typical hydrogen diffusion reaction of process {\bf 2.}. A hydrogen is released from 
an Si-O-H (O$^*$) to another Si-O-H group (O$^*$) forming a water like structure. An Si-O dangling bond and 
an relaesed hydrogen atom are the resulting products. }
\end{figure}

In this section we discuss the  possible mechanisms for hydrogen diffusion as it has been mentioned 
in the introduction. We emphasize again that, due to the presence of thermostats, dynamical quantities like the 
diffusion constants cannot be extracted reliably.
Nevertheless, the structural characteristics of the melt should allow to obtain at least some insight into its
dynamical properties. 
Our aim is therefore to determine whether the structural units we discussed in Sec. \protect\ref{sec:FND},
such as the O3H clusters or the Si-O dangling bonds, 
can serve as intermediate states for the hydrogen diffusion processes in SiO$_2$-H$_2$O liquids.  \\
We start this discussion by making a list of the possible hydrogen diffusion mechanisms in liquid
silica, eliminating the free water molecules or stable O-H groups as possible free hydrogen carriers
since, as discussed above, they are absent in our simulation data. 
As the hydrogen atoms are attached to the silica matrix 
in the form of Si-O-H groups, three possible mechanisms come into play:
\begin{description}
\item[{  1.}] motion of the hydrogen in the form {\rm O-H- - -O $\longrightarrow$ O- - -H-O}
\item[{  2.}] motion of the oxygen in the form {\rm H-O- - -H $\longrightarrow$ H- - -O-H}
\item[{  3.}] motion of the O-H group in the form {\rm Si-(OH)- - -Si $\longrightarrow$ Si- - -(OH)-Si.}
\end{description}
The first two  mechanisms require the rupture of an O-H bond whereas the third one requires the
rupture of a Si-O bond. 
A typical reaction involving the two first processes are shown in Figs. \ref{fig:diffreact1} and \ref{fig:diffreact2} .

In the simulated trajectories, we can easily count the number of O-H ruptures in order to 
find out whether the first two  mechanisms exist in the liquid. However, it is possible to
distingish the two processes only by looking at the reactants and the decay products associated to each
encountered O-H rupture. 
The possible reactants (resp. decay products) for the formation (resp. rupture) of an O-H group are given in Tab.
\protect\ref{tab:decay} where the O$^*$ refers to an oxygen in a Si-O-H group, wO to an oxygen atom in a water molecule,
BO to an oxygen atom in a Si-O-Si group and O3H to an oxgen atom in a Si-(OH)-Si group (see the
definitions in the introduction of Sec. \protect\ref{sec:res}). Note that the formation or rupture of an O-H bond associated 
with a wO $\longrightarrow$ O$^*$ requires the water molecule to be close to a silicon atom which is 
indeed the case since we have deduced from Fig. \protect\ref{fig:coordtemp} that no free water molecule exist
in the liquid.
\begin{table}[h]
\begin{center} 
\begin{tabular}{ll||ll}
\hline \hline 
\multicolumn{2}{c||}{{\sf O-H formation}} & \multicolumn{2}{c}{{\sf O-H rupture}} \\
{\sf reactant} &  {\sf final state\phantom{a}}  & {\sf initial state} &  {\sf decay product} \\ \hline
Si-O dangling & O$^*$ & O$^*$ &  Si-O dangling \\
O$^*$ & wO  & wO &  O$^*$  \\
BO & O3H  & O3H &  BO   \\ 
\hline \hline
\end{tabular} 
\end{center}

\caption{\label{tab:decay} List of possible reactants for the O-H formation and of possible decay products for the O-H rupture.
The notations refer to the definition of the different oxygen types given in the introduction of Sec. \protect\ref{sec:res}. The final (resp. initial) state gives the type of the oxygen atom in the formed (resp. destroyed) O-H unit.}
\end{table}

\begin{figure}[h] 
\centerline{\includegraphics[width=5cm,angle=-90]{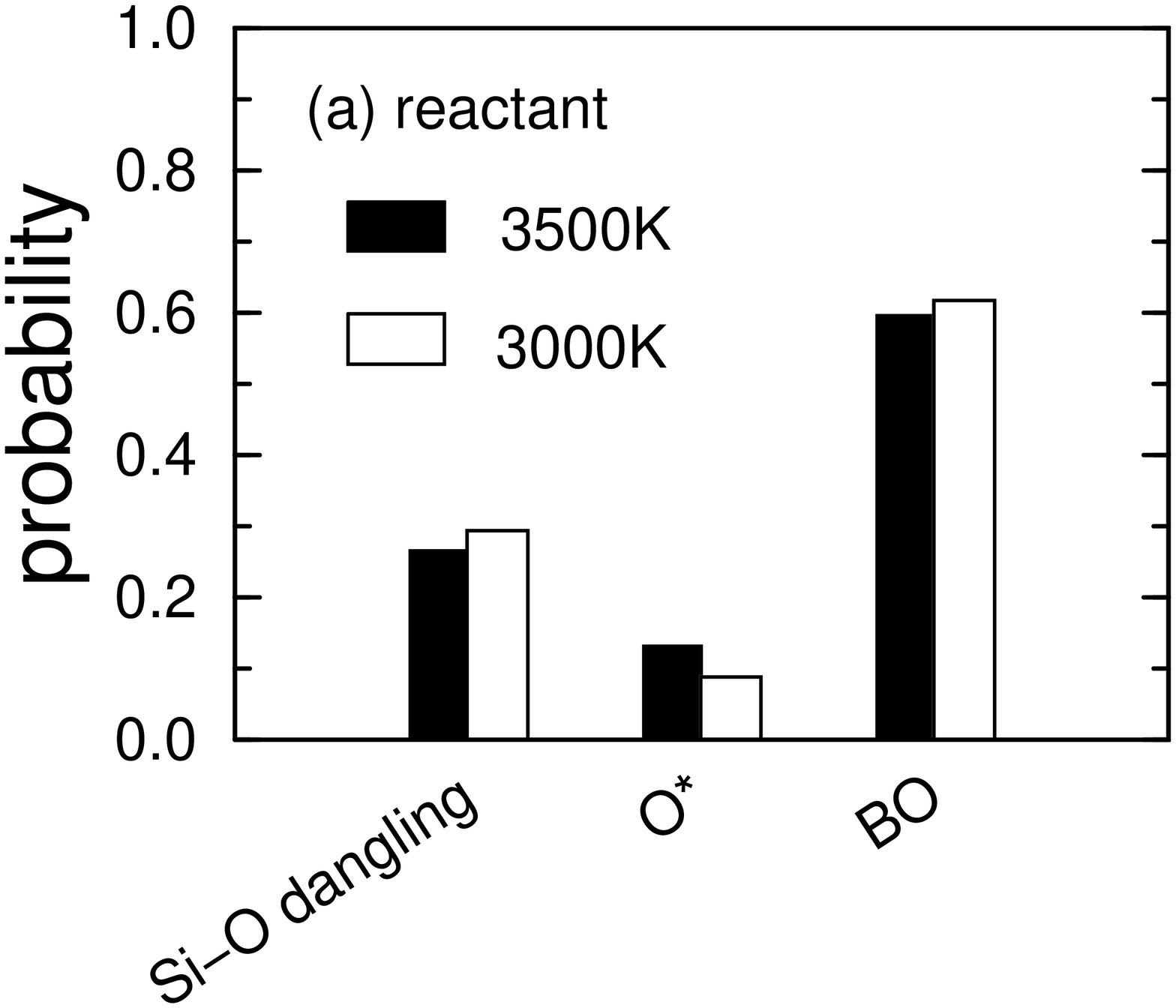}
\includegraphics[width=5cm,angle=-90]{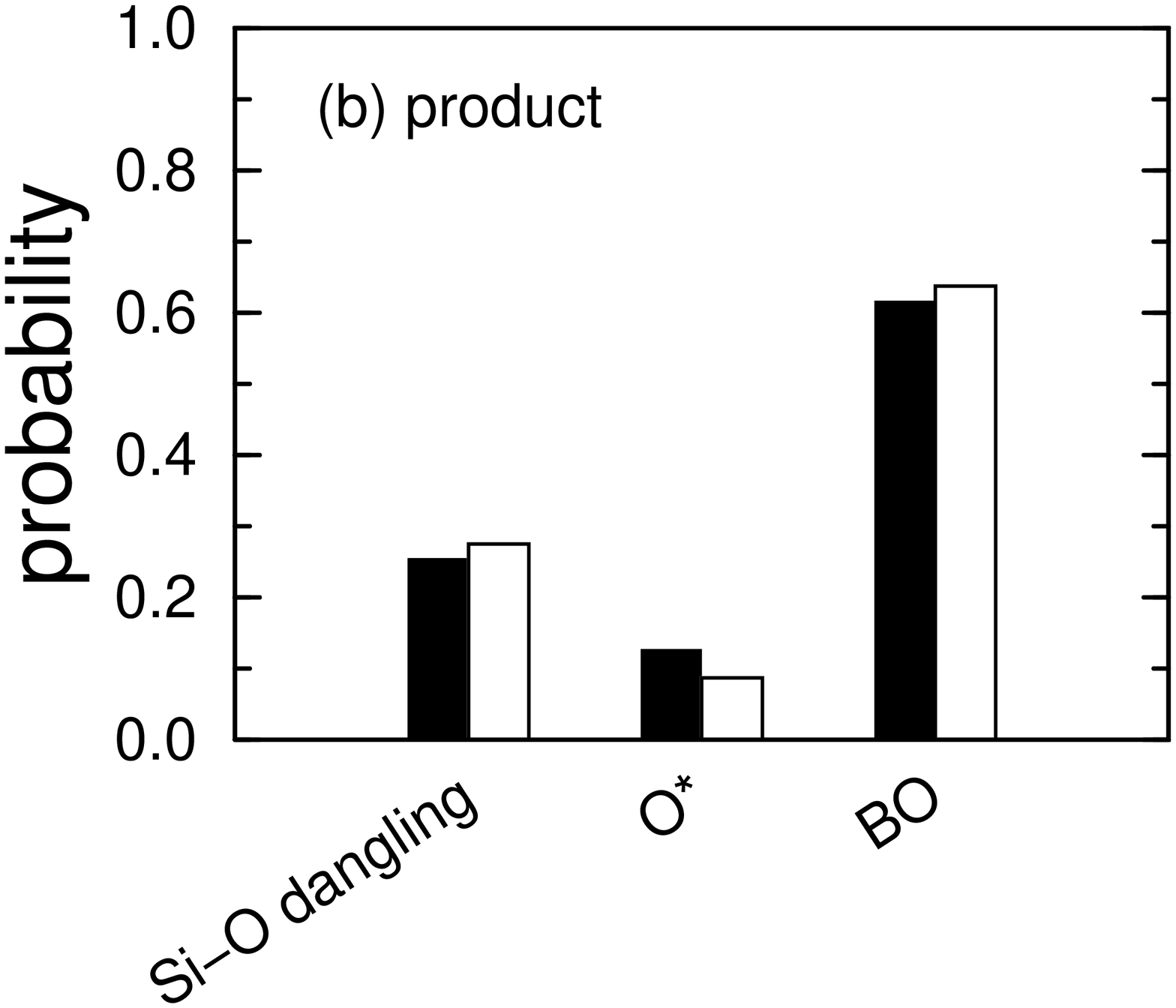}}

\vspace{0.5cm}
\centerline{\includegraphics[width=5cm,angle=-90]{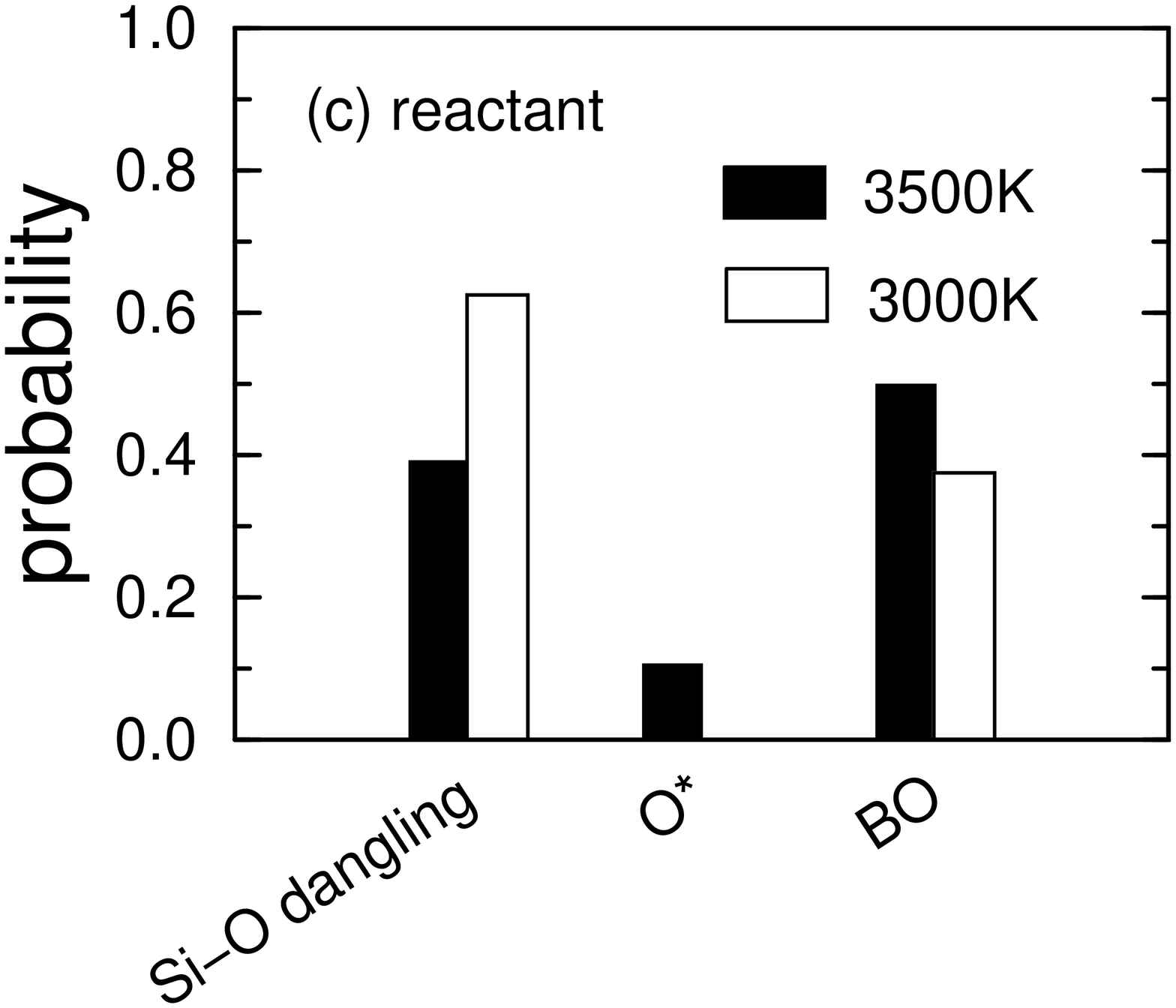}
\includegraphics[width=5cm,angle=-90]{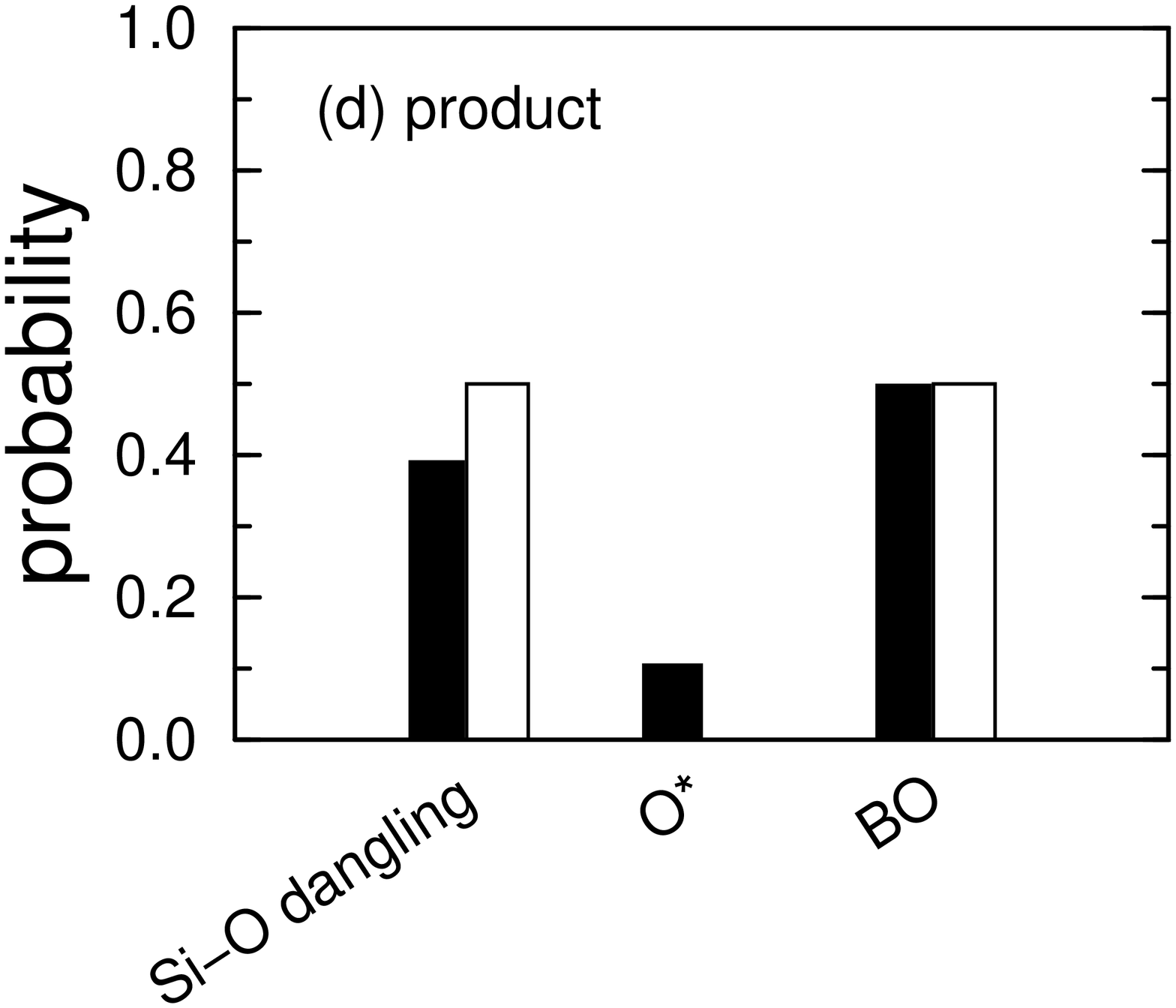}   } 
\caption{\label{fig:decOHtrans} Probabilities of O-H group formation and decay including recombinations, (a) and (b), 
and without recombinations, (c) and (d),  associated to Si-O dangling bonds, O$^*$ and BO. 
Black and white bars correspond to 3500 K  and  3000 K, respectively.}
\end{figure}

The average number of O-H bonds in the SiO$_2$-H$_2$O melts
is found to be equal to 8.4 at 3500 K and to 8.1 at 3000 K and is therefore larger than the total number of 
hydrogen atoms in the system. Hence we conclude that intermediate states with two oxygen atoms close to a single 
hydrogen exist and that they must serve as intermediate states for hydrogen exchange reactions (process number {\bf 1.}). \\
By counting the number of O-H ruptures along the
trajectories and the type of reactants/decay products associated to these ruptures, we are able to show the existence of hydrogen diffusion processes of types {\bf 1.} and {\bf 2.}. 
 For the following we assumed that an O-H bond was formed if the O-H interatomic distance was 
larger than the O-H cutoff at step $t$ and smaller than the O-H cutoff at step $t+\Delta t$ ($\Delta t$ being the length of a time step of 0.1088 fs). The nature of the reactant is found by counting, at time $t$, all silicon and hydrogen neighbors (again in terms of the nearest neighbor cutoff) of the later O-H oxygen.
A rupture of an O-H bond was assumed to have happened 
if one H and one O atom had an interatomic distance smaller than the O-H
cutoff at time $t$ and if their interatomic distance 
 became larger than the O-H cutoff at step $t+\Delta t$, irrespective of whether subsequently the same
bond was formed again or not. The decay product
of an O-H rupture is the H donator unit (without the H atom) and is
found by counting, at step $t+\Delta t$, all silicon and hydrogen neighbors (again in terms of
the nearest neighbor cutoff) of the former O-H oxygen. \\
At 3500 K the total number of O-H formations and ruptures was 97 and 94 respectively,
 during the equilibrated part of the trajectory (8.1 ps). 
At 3000 K we found 68 O-H formations and 69 ruptures during the equilibrated part of the trajectory (11.6 ps). 
However, in order to take into account the O-H formations and ruptures that serve for the hydrogen diffusion, we
also have to distinguish between the formations and ruptures that give rise to {\it hydrogen transfers} or 
{\it recombinations}. In order to perform this separation between transfers and recombinations it is necessary to relate each decay of an OH bond to a formation of an OH bond. A hydrogen transfer requires the rupture of one O-H bond 
and the formation of a {\it different} one, 
whereas a recombination implies only one bond. Note that the rupture of the first bond will occcur before the formation of the new bond if a transient free hydrogen is formed, or, as in almost every case, after the formation of a new bond which implies the formation of the intermediate state.  Making this distinction between transfers and recombinations, 
we find 28 transfers at 3500 K (corresponding to a ratio of $\frac{28}{94}=28.8 \ \%$ transfers 
and 71.2 $ \%$ recombinations) and 8 transfers at 3000 K (corresponding to a ratio of $\frac{8}{68}=11.8 \ \%$
 transfers and $88.2 \ \%$ recombinations).  The ratio between  
the number of ruptures involved in a transfer and the total number of ruptures gives the recombination rate.\\
Figs. \ref{fig:decOHtrans}a and b give the percentage of the reactants and decay products for the O-H formations
and ruptures including recombinations 
and Figs. \ref{fig:decOHtrans}c and d the related transfer reactions (i.e. without recombinations).
For the transfers, due to the small numbers, the error bars are quite large. 
At both temperatures, the contributions of the Si-O dangling bonds are higher 
for the transfer related ruptures than for the overall ruptures, i.e. transfers and recombinations. 
Furthermore we recognize from Figs. \ref{fig:decOHtrans}c and d that the 
Si-O dangling bonds and the BOs are the main acceptors and rupture products, with almost equal probability. 

Concerning transfers, the Si-O dangling bonds react with a probability of 40 \% at 3500~K and 60 \% at 3000 K 
into an SiOH group and 
the SiOH group (O$^*$) decays with a probability of 40 \% at 3500 K and 50 \% at 3000 K into a Si-O dangling bond. 
However the corresponding probability that includes recombinations is for both the formation and the decay around 30 \%. This result indicates that  the Si-O dangling bonds are more involved in the formation and rupture
of the O-H bonds when the transfers are considered and the recombinations are taken out. Furthermore it also 
demonstrates the existence of large vibrations of the H atoms around the O atoms of the Si-O dangling bonds 
and hence the existence of the "weak" hydrogen bonds described in Sec. \protect\ref{sec:RDF}. \\
The BOs serve, with a probability of 50 \% at 3500 K and of 40 \% at 3000 K, as reactants for
 a hydrogen transfer leading to a new O-H group (60 \% including recombinations at both temperatures). 
On the other side, the BOs serve as products of a hydrogen transfer with a probability of about 50 \%, and
therefore  the participation of the BOs to the formation and rupture of the O-H bonds is slightly decreased when transfers are compared to the reactions including recombinations. \\
Finally, the O$^*$ contribute much less to the O-H group formation and decay 
than the other species, with a probability 
of about 10 \% (12 \% including recombinations) at 3500 K and 0 \% (8 \% including recombinations) at 3000 K.\\
From Fig. \protect\ref{fig:decOHtrans}, we deduce that hydrogen transfers involving  Si-O dangling
bonds or BOs as reactants as well as decay products dominate with $\approx$ 90 \%. Since these units
are associated to O$^*$ or O3H as initial or final states, they correspond to the hydrogen diffusion
process number {\bf 1.} which means that among the processes {\bf 1.} and {\bf 2.}, the former dominates
with $\approx$ 90 \%.

\begin{figure}[h] 
\centerline{\includegraphics[width=14cm]{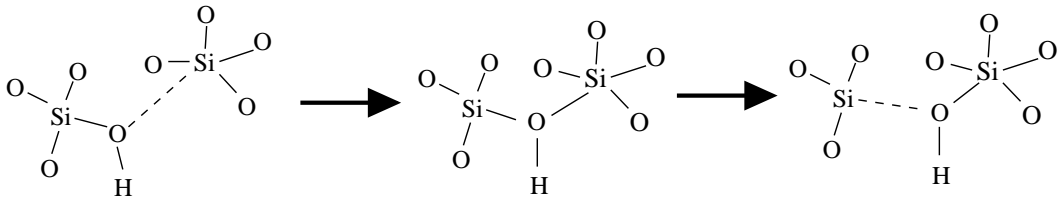}  }
\caption{\label{fig:diffreact3} Typical hydrogen diffusion reaction of process three involving a O3H. 
A O-H group is released from an Si-O-H (O$^*$) to a O3H cluster decaying subsequently 
in another O$^*$. An oversaturated  and 
an unsaturated silicon atoms are subsequent products. }
\end{figure}

Let us now consider the third considered hydrogen diffusion mechanism. It is associated to the motion
of a O-H unit as a whole (without any O-H rupture) which can be obtained via the formation
and decay of an O3H cluster as sketched in Fig. \protect\ref{fig:diffreact3}. 
Note that this process requires the presence  of "weak" Si-O bonds, the existence of which  was inferred 
from Fig. \protect\ref{fig:specOSi} and leads to the formation of threefold and fivefold coordinated silicon atoms. 
As for the O-H bonds, we  have analyzed the probabilities to find the different reactants and decay products 
for the O3H clusters formation and rupture. 
Along the equilibrated part of the trajectories, we have found 142 formations and 139 ruptures at 3500 K and 
104 formations and 103 ruptures at 3000 K. In the case of O3H clusters, the possible reactants and decay
products are the oxygen atoms involved in a Si-O-H unit (O$^*$) or a bridging oxygen (BO). The latter case involves
an O-H rupture event whereas the first case does not. This is the reason why  
a distinction between {\it recombinations} and {\it transfers} is not easily feasible for these species since 
the transfered object can either be a hydrogen atom (as in Tab. \protect\ref{tab:decay}) or an entire O-H group (as in Fig. \protect\ref{fig:diffreact3}).
We hence display the distribution for O3H formations and ruptures including recombinations 
\begin{figure}[h] 
\centerline{\includegraphics[width=4cm,angle=-90]{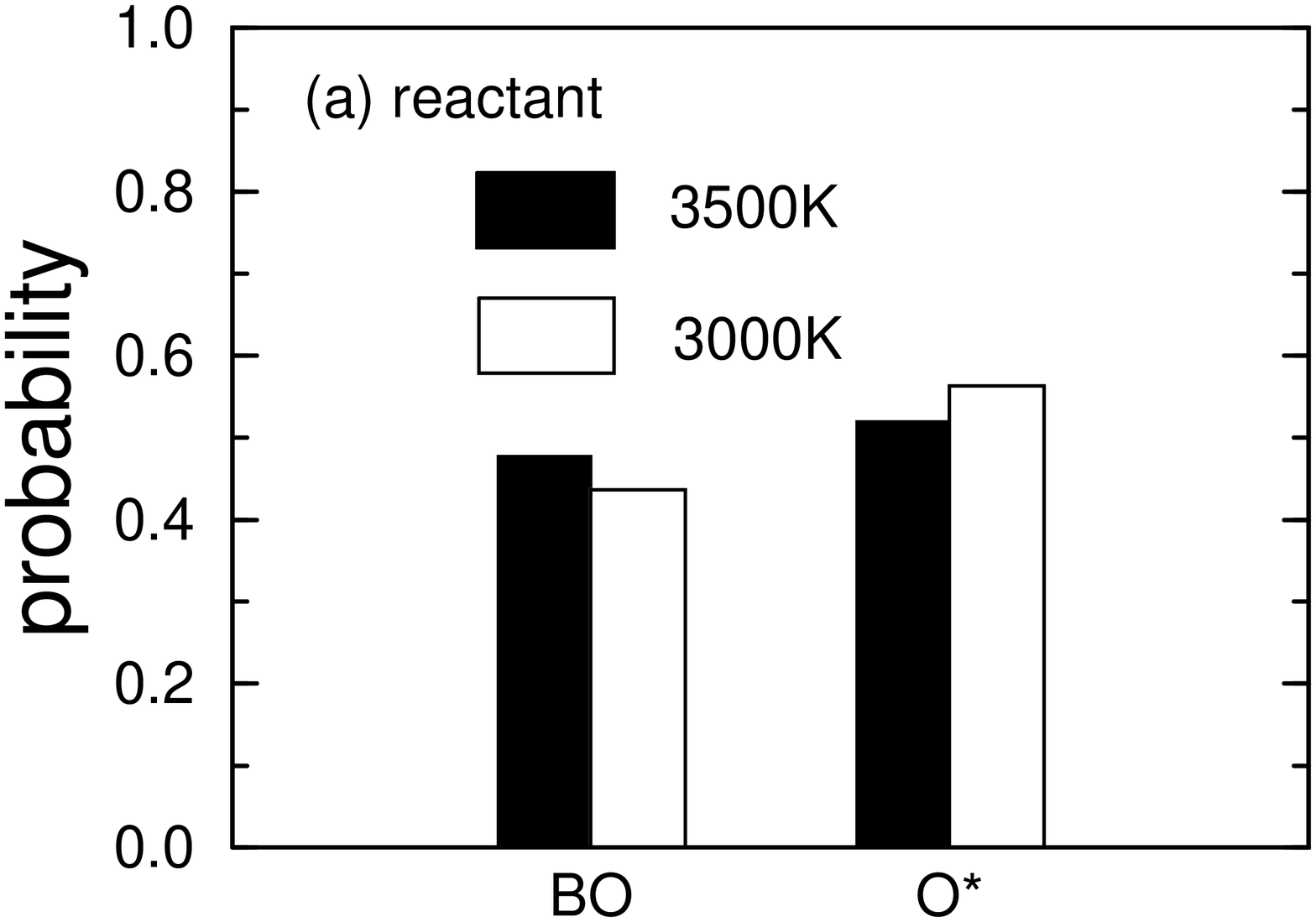}
\includegraphics[width=4cm,angle=-90]{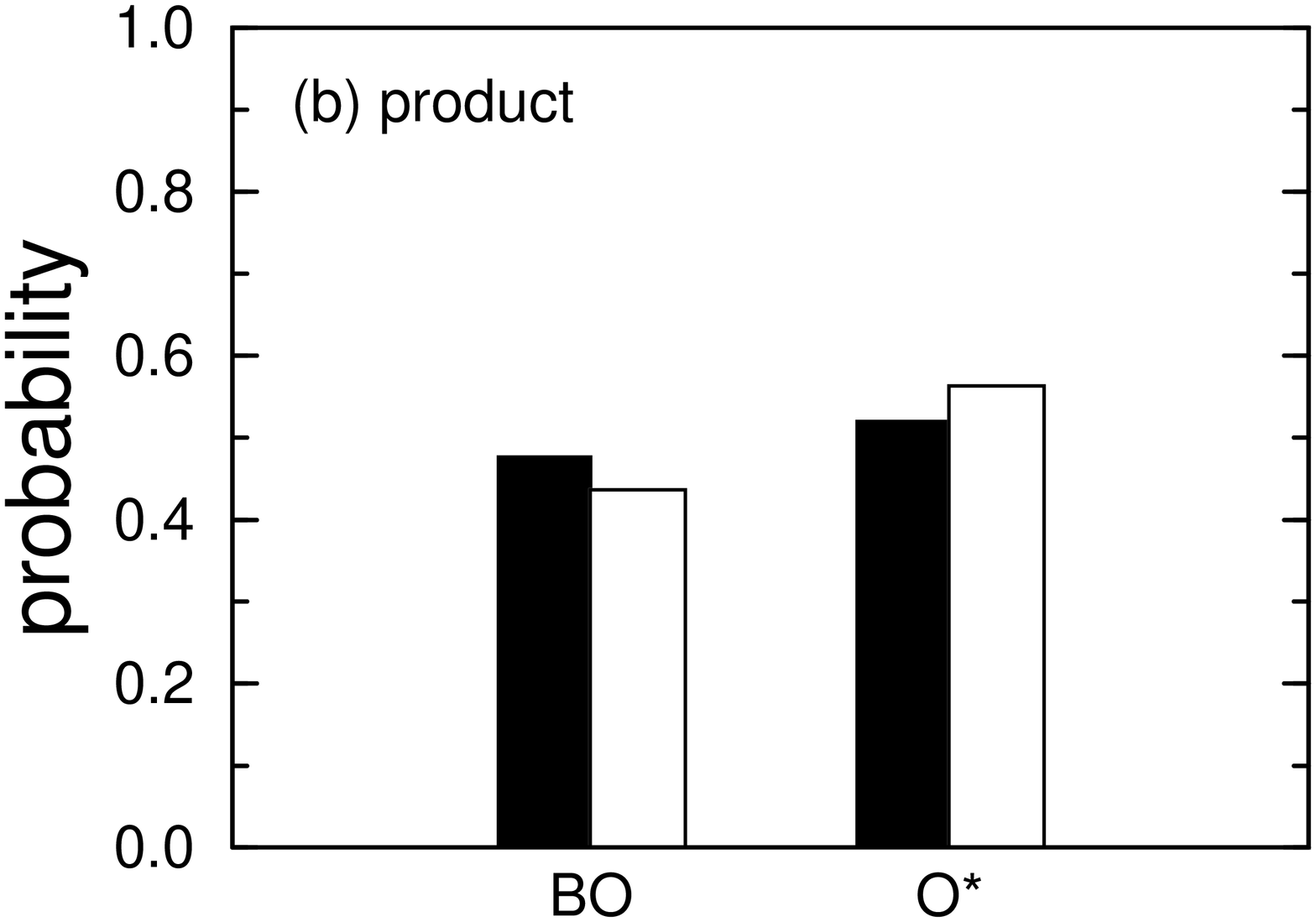}  }
\caption{\label{fig:decO3H} O3H group formation (left) and decay (right) 
 products at 3500 K (black bars) and 3000 K (white bars)
including recombinations.}
\end{figure}
in Fig. \ref{fig:decO3H}. There are only very small differences between formation and decay.
At both temperatures, the O3H clusters form and decay from and into BO and O$^*$, with almost the same probabilities.  
However at 3000 K, the probability from and into O$^*$ is slightly preferred with a value of 56 \% for O$^*$  vs. 44 \%
for BO. This indicates that the hydrogen diffusion process involving the motion of a O-H group without any
O-H bond rupture (process number {\bf 3.}) occurs  in the liquids as well.

To conclude the section on the diffusion process, we analyzed the activation energy for a hydrogen transfer
corresponding to the rupture of an O-H bond (without recombination). 
We can calculate the activation energy for a hyrogen transfer reaction assuming an Arrhenius behavior
\begin{equation}
R(T)=A\:{ \exp }\left\{-\frac{E_A}{k_BT}\right\} \ ,
\end{equation}
where the reaction rate $R(T)$ is given by a constant $A$ and the activation energy $E_A$. Using 
the two numbers of O-H ruptures leading to transfers of 28 for 3500 K and 8 for 3000 K we obtain 280 kJ/mol (2.91 eV) 
for the activation energy (the different lengths of the equilibrated trajectories already taken into account). 
Unfortunately the constant $A$ depends explicitly on the transfers per time and can hence not be determined due 
to the above mentioned problem of the thermostats. The activation energies for the generation and dissociation of various Si$_x$O$_y$H$_z$ molecules in the gas phase have been determined by Zachariah {\it et al} \cite{Za95} with the GAUSSIAN {\it ab initio} package \cite{Ga90}. Zachariah {\it et al} found a value for the activation energy of the formation of OSiOH from SiO$_2$ and H of 285 kJ/mol which compares very well to the above given one of 280 kJ/mol for a hydrogen transfer.

\section{Summary}
We have equilibrated a liquid of silica containing  3.84 wt. \% H$_2$O at 3500 K and 3000 K using first-principles 
molecular-dynamics simulations. We observed that the water is mostly attached to the silica network in the form of Si-OH groups. 
Water molecules or free O-H groups occur only at the highest temperature but are not stable and decay rapidly. 
The SiO$_4$ tetrahedra are still the basic network forming units, as in pure silica. 
The short range correlation (i.e. the radial  distribution functions) suggest that the structure of the matrix is as much changed by the addition of water than as by the addition of the 
same amount of sodium oxide to the liquid. In contrast to this, the way the modifier cation itself is attached to the matrix seems to be quite different. The sodium atoms tend to form bonds of ionic character with oxygen, whereas the hydrogen atoms are attached by covalent bonds. 
This difference in the bonding character of  O-H and O-Na could be the reason for the slower diffusion of the hydrogen atoms in liquid silica compared to that of the sodium atoms. \\
The neutron scattering structure factor shows a pronounced prepeak
 at a wave vector of about 1\AA $^{-1}$, i.e. we have evidence for the presence of long range correlations. 
The prepeak is even more intense than that for the sodium silicate at the 
same wave vector transfer. The origin of this feature in the hydrous liquid seems to be different from that of the 
sodium silicate, since the partial structure factor for H-H contributes only little. Furthermore, in the hydrous 
case, the silica matrix itself seems to be modified since the prepeak occurs as well in the partial structure factors 
for the matrix. It remains to be investigated whether the prepeak can be observed at ambient temperature. The  data 
show a shift to higher $q$-vectors and a decrease of the intensity at the lowest temperature. \\ 
The other experimentally accessible quantity, the Q$^n$ distribution, shows clearly that the O-H groups try to avoid each other. Silicon atoms with two and more Si-O dangling bonds or O$^*$ (Q$^n$ sites with $n\leq$ 2) are rarely found.
Despite the strong covalent character of the O-H bond, the O-H units have a high radial mobility and oxygen triclusters have
a high tendency to form. These oxygen triclusters violate the oxygen stoichiometry and the system 
compensates this violation by the formation of Si-O dangling bonds. \\
The dangling bonds and the triclusters constitute mainly the intermediate states for the 
diffusion mechanism of the hydrogen atoms in the melt. 
Indeed three possible diffusion mechanisms have been discussed, two of them involving the rupture of an O-H bond. 
Free water molecules, hydrogen molecules and free hydroxyl groups are not or only rarely found and cannot therefore 
be made responsible for the hydrogen diffusion in the liquid. \\      
Coming back to one of the motivating questions, the bubble formation, only little can be concluded from this study. 
On the one hand, it turned out that the O-H groups are very stable species and it requires a certain stoichiometry
 violation to weaken the O-H bonds (note that in O-H-O transition units the stoichiometry of the H atom is also violated). 
On the other hand these bond weakening transition states occur in sufficiently high quantity to assure the hydrogen diffusion. 
Nevertheless, a clustering of hydrogen and the formation of stable water molecules do not take place, especially at the lower 
temperature. If a the  formation of bubbles were realized, we would expect a higher signal in the H-H radial distribution function at typical H-H distances found
 in liquid water (2-3 \AA ) and perhaps a signal in the Q$^n$ distribution for $n\leq$ 2. 
Hence we do not see evidence for water clustering at the considered temperatures. 
We recognize that the influence of temperature, pressure and silicate composition on the bubble 
formation is not yet investigated and further work of the present kind will be mandatory.

\section{Acknowledgments}
M. P. thanks the Bund der Freunde der Technischen Universit\"at M\"unchen and the French Science Ministry for a grant. 
The authors thank also Andreas Meyer and Winfried Petry from the Technische Universit\"at M\"unchen and Helmut Schober and Mark Johnson from the Insitut Laue-Langevin for fruitful discussions and Simona Ispas for the NS4 liquid simulation data.
The simulations have been performed on the Hitachi SR8000 at Leibniz Rechenzentrum, Munich, Germany 
and on the IBM/SP3 at CINES, Montpellier, France.

\end{document}